\documentclass{ws-ijbc}
\usepackage{hyperref}

\begin{document}

\title{A Dynamical Model of Decision-Making Behaviour in a Network of Consumers with Applications to Energy Choices.\footnote{Electronic version of an article published in International Journal of Bifurcation and Chaos 21, 2467 (2011). DOI: \href{http://www.worldscientific.com/doi/abs/10.1142/S0218127411030076}{10.1142/S0218127411030076}  \copyright World Scientific Publishing Company \url{http://www.worldscientific.com/worldscinet/ijbc}}}

\author{N.\ J.\ McCullen\footnote{Now at the University of Bath. Permanent email: n.mccullen@physics.org}}

\address{School of Mathematics, University of Leeds\\
Leeds, LS2 9JT, United Kingdom}

\author{M.\ V.\ Ivanchenko\footnote{Currently at University of Nizhny Novgorod, Russia}}
\address{School of Mathematics, University of Leeds\\
Leeds, LS2 9JT, United Kingdom}

\author{V.\ D.\ Shalfeev}

\address{Department of Radiophysics, Nizhny Novgorod University\\
23 Gagarin Avenue, 603600 Nizhny Novgorod, Russia\\
shalfeev@rf.unn.ru}

\author{W.\ F.\ Gale}

\address{School of Process, Environmental and Materials Engineering, University of Leeds\\
Leeds, LS2 9JT, United Kingdom.}

\maketitle

\begin{abstract}
A consumer Behaviour model is considered in the context of a network of interacting individuals in an energy market.
We propose and analyse a simple dynamical model of an ensemble of coupled active elements mimicking consumers' Behaviour, where ``word-of-mouth'' interactions between individuals is important. 
A single element is modelled using the automatic control system framework. 
Assuming local (nearest neighbour) coupling we study the evolution of chains and lattices of the model consumers on variation of the coupling strength and initial conditions. 
The results are interpreted as the dynamics of the decision-making process by the energy-market consumers.
We demonstrate that a pitchfork bifurcation to the homogeneous solution leads to bistability of stationary regimes, while the autonomous system is always monostable. 
In presence of inhomogeneities this results in the formation of clusters of sharply positive and negative opinions. 
We also find that, depending on the coupling strength, the perturbations caused by inhomogeneities can be exponentially Localised in space or de-Localised. 
In the latter case the coarse-graining of opinion clusters occurs. 
\end{abstract}


\section{Introduction}

The Behaviour of ensembles and networks of coupled active elements, or oscillators, has long been the focus of attention of research into the dynamics of complex systems \cite{haken1977synergy, golubitsky2004some}. 
Examples include communication networks of coupled oscillators, electrical generation networks, biological and artificial neural networks \cite{abarbanel1996synchronisation, watts1998collective, strogatz2001exploring}. 
Not unexpectedly, there has also been interest in modelling socio-economic systems \cite{gaertner1974dynamic, bass2004comments, weidlich2003sociodynamics}. 
The nature of such systems, however, dictates that the models' parameters are hard to measure or estimate, if indeed it is possible at all. 
Nevertheless, the analysis of these models uncovers qualitative characteristics and dynamical trends in Behaviour, such as potential scenarios for the evolution of socio-economic systems \cite{bak1996nature}. 
Analysing opinion-formation in social systems belongs to this class of problems \cite{stauffer2005sociophysics}.

In this paper we study a dynamical model that can be used to describe the Behaviour of an ensemble of consumers, in particular in the energy market. 
Specific interest lies in understanding the potential regimes of collective Behaviour and responses of the end-users or mediators, making decisions that are influenced by the ideas and actions of peers.
The underlying intent of the model is to inform the thinking of policy makers who are seeking to evaluate the potential effectiveness of interventions to promote energy sustainability.
To model the consumers we use the framework of automatic control systems \cite{kuo1981automatic}, where individuals attempt to regulate some variable to match a reference value.
This was chosen as a way to formulate the model system as the primary motivations in the decision-making process can be clearly related to response functions.
It also has the advantage of making few and simple assumptions about the Behaviour of the individuals, whilst retaining the essential features of  the Behavioural responses to the external stimuli.

We start with a simple model characterized by a single variable, namely the price of a commodity which is specific for each consumer and the market in general. 
The commodity in question could be from a range of things, such as energy purchased from a specific supplier, or a technology for generating or saving energy in the home such as the installation of photovoltaic panels or replacing an appliance with a more energy efficient model.
Making the decision whether to buy (or continue buying), the consumers compare the market price with what they consider some ``reasonable'' price in their own mind.
This so called ``fair'' price is formed with reference to the consumers' own previous estimate as compared to the current market price, as well as the information coming from the opinions of the other consumers.
This leads to the formulation of a dynamical network model of the consumers' decision-making on the market.
The automatic control system with a feedback that we use to model the decision-maker conveniently embodies the above properties.
For the sake of simplicity we restrict our analysis to one-dimensional chain and two-dimensional lattice topologies with only nearest-neighbour coupling. 

\section{Basic model}

\begin{figure}[h!!]
\centering
\epsfig{file=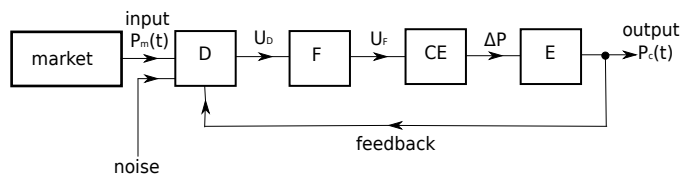,width=0.5\linewidth}
\caption{Model of the consumer (MC) as an automatic control system.}\label{fig:controlsystem}
\end{figure}

The basic model of a consumer (MC) as an automatic control system is shown in Figure \ref{fig:controlsystem}.
The object under control is known as the estimator (E), which gives the consumer's current estimate of the reasonable price of the goods as the output signal $P_c(t)$.
The market price of the goods (taken as constant in the decision time-scale) is the input reference signal $P_m(t)$, which is compared to the consumer's own fair-price, $P_c(t)$.

The current output signal $P_c(t)$ and the input reference signal  $P_m(t)$ are compared using the discriminator function (D).
The output signal of D ($U_D$) is passed through a filter (F), which eliminates noise parasitical components of upper frequencies from the spectrum (as the customer would average the short time-scale price fluctuations on the market).
Then the signal from the F output ($U_F$) is supplied to the control element (CE), which directly changes the consumer's estimated price in a manner such as to approach the reference market price.
This is interpreted as the consumer changing their opinion of the fair-price over time in response to the current market price.

In order to obtain the equations describing the dynamics of such an automatic control system, take the equations for each element of the system in turn.
The equation for the output of the estimator E can be written as:
\begin{equation}
P_c=(P_c)_i+\Delta P,\label{eqn:E}
\end{equation}
where $(P_c)_i$ is the consumer's initial estimate of the fair price before the feedback comparison to the market price is considered, $\Delta P$ is the change in the estimated price as controlled by CE.

\begin{figure}[h!!]
\centering
\epsfig{file=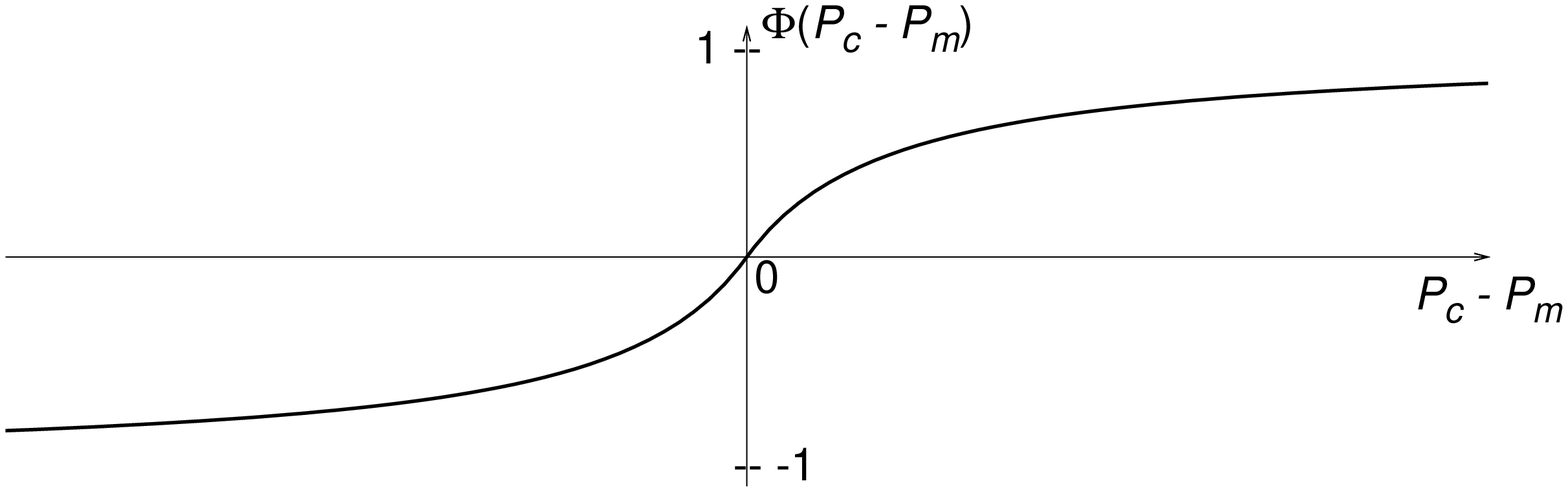,width=0.5\linewidth}
\caption{The nonlinear characteristics of the discriminator function.  In this example the impact of extremely high and low deviations does not increase proportionally, perhaps due to being seen as unrealistic by the consumer.}\label{fig:discriminator}
\end{figure}

The equation for the control element CE can be written as:
\begin{equation}
\Delta P = -S U_F,\label{eqn:CE}
\end{equation}
where $S$ is the slope of the CE characteristics and the minus sign means that the action of control is to bring $P_c$ towards $P_m$.
The equation for F can be written as:
\begin{equation}
U_F = K(p) U_D,\ \mathrm{with}\ p\equiv\frac{d}{dt},\label{eqn:F}
\end{equation}
where $K(p)$ is the filter transmission factor. 
The equation for the discriminator D can be written as:
\begin{equation}
U_D = E \Phi(P_c-P_m),\label{eqn:D}
\end{equation}
where $E$ is the maximum output of D and $\Phi(P_c-P_m)$ is the discriminator's nonlinear characteristics, normalized to unity (Fig.~\ref{fig:discriminator}).

Introducing the current price deviation $P(t)=P_c(t)-P_m(t)$, the parameter $\sigma=SE$ as the greatest error to be corrected by the control circuit, dimensionless price-deviation $X=\dfrac{P}{\sigma}$ and dimensionless initial price-deviation $\gamma=\dfrac{P_i}{\sigma}$, we obtain from (\ref{eqn:E})--(\ref{eqn:D}) the following equation, describing the dynamics of the model of the consumer (MC) as an automatic control  system:
\begin{equation}
X + K(p) \Phi(X) = \gamma,\ p\equiv\frac{d}{dt}.\label{eqn:MC}
\end{equation}
The initial condition is also specified here, defined in (\ref{eqn:MC}) as $X(0)=\gamma$. 

The model (\ref{eqn:MC}) addressed here looks like the standard continuous model of automatic synchronization systems widely applied in synchronization theory \cite{afraimovich1994stability} and oscillation theory \cite{andronov198chaikin}.

\section{Single Consumer}

In the simplest case of an integrating filter $K(p)=\dfrac{1}{1+ap}$, introducing dimensionless time $\tau=\dfrac{t}{a}$ and assuming $\gamma$ remains constant (meaning the market price $P_m=\mathrm{const}$); instead of (\ref{eqn:MC}) we have a differential equation of first order:
\begin{equation}
\frac{dX}{d\tau} + X + \Phi(X) = \gamma.\label{eqn:ode}
\end{equation}

In equation (\ref{eqn:ode}), for a symmetric nonlinearity $\Phi(X)$ (as in Fig.~\ref{fig:discriminator}) it suffices to treat the parameter $\gamma$ as non-negative, because for $\gamma < 0$ we can change the sign of $X$ to $-X$ and get the same equation (\ref{eqn:ode}).
The coordinate of the equilibrium states of equation (\ref{eqn:ode}) can be found from the equation:
\begin{equation}
\gamma-X=\Phi(X).\label{eqn:balance}
\end{equation}

\begin{figure}[!!h]
\centering
\epsfig{file=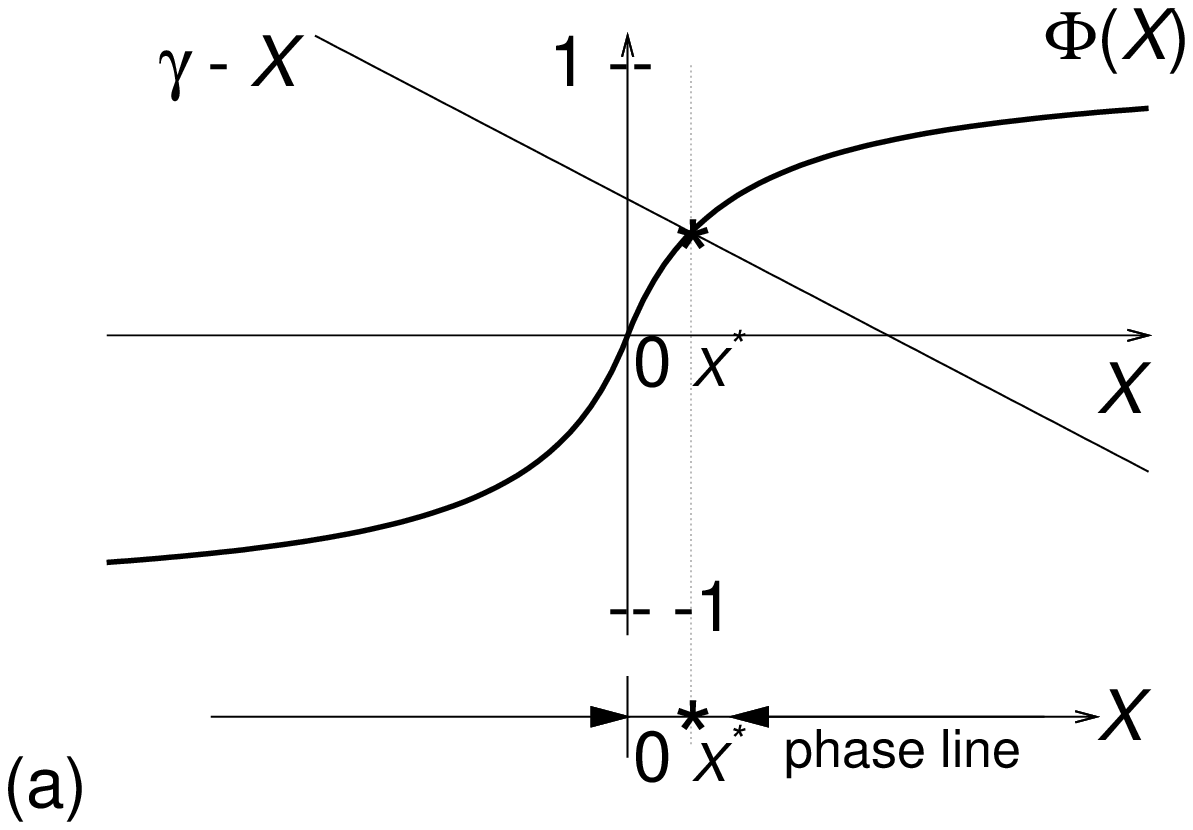,width=0.3\linewidth}
\epsfig{file=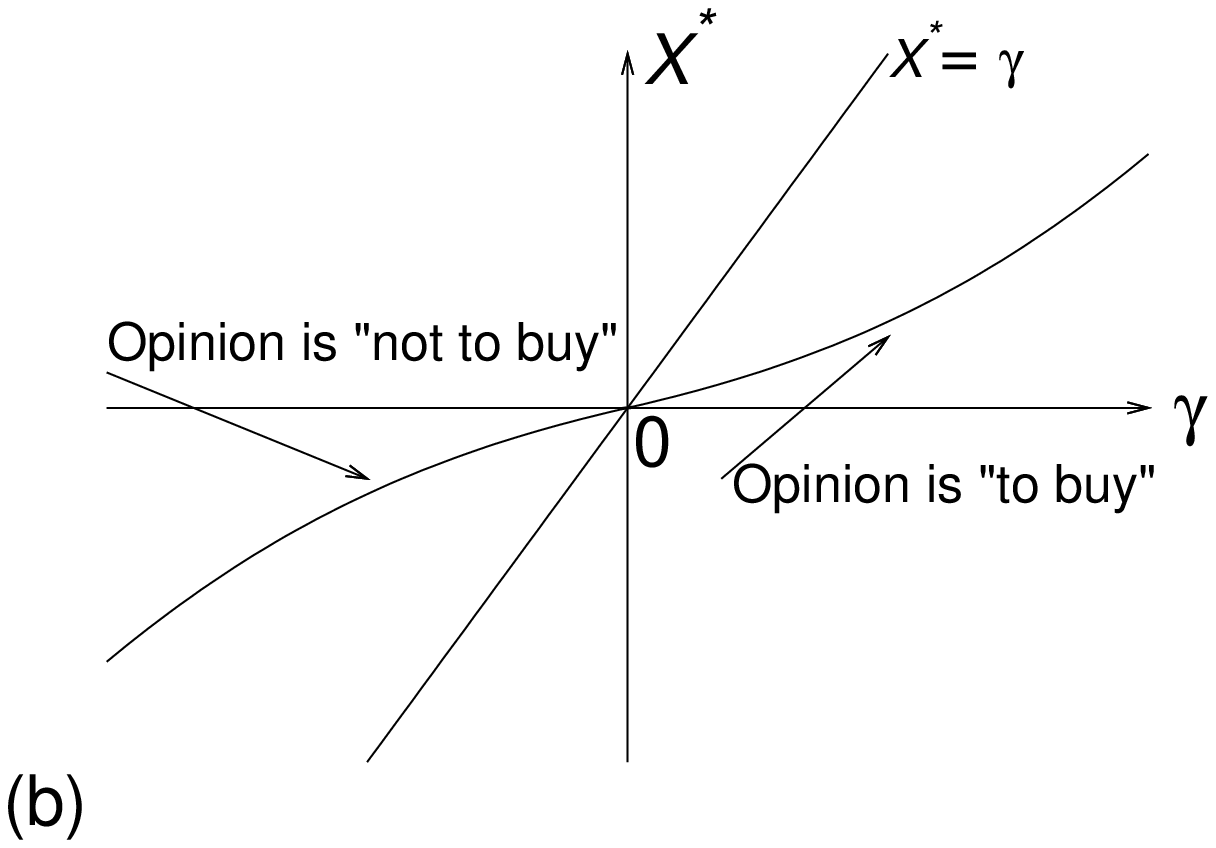,width=0.3\linewidth}
\caption{Phase portrait of (\ref{eqn:balance}), showing how the initial estimate $\gamma$ and nonlinear function $\Phi(X)$ determine the final opinion $X^*$, and the relationship $X^*(\gamma)$, with the corresponding decision by the consumer.}\label{fig:phase}
\end{figure}

According to equation(\ref{eqn:balance}) there is only one stable equilibrium state with coordinate $X^*$ on the phase line $X$ (Fig.~\ref{fig:phase}(a)).
Figure \ref{fig:phase}(b) illustrates the dependence $X^*(\gamma)$, the final opinion state arising from the initial estimates (i.e.\ the resulting decision).
In all cases $|X^*| \leq |\gamma|$. 
Thus in the case of a single consumer the model (\ref{eqn:ode}) demonstrates very simple dynamics and such Behaviour looks reasonable.
If the consumer's initial estimate of the price deviation $\gamma > 0$ then, after comparing the initial and market prices, the resulting estimate of the difference between the market price and the fair price by the consumer becomes $0 < X^* < \gamma$.
This means that the consumer's final opinion is positive (being proportional to $X^*$) and the consumer makes the decision ``to buy'' (Fig.~\ref{fig:phase}(b)).
In contrast, if the initial estimate is that the market price is higher than what they consider ``fair'' ($\gamma < 0$), the resulting estimate becomes $X^* < 0$, and $|X^*| < |\gamma|$.
This means that the consumer's final opinion is negative and the  decision is ``not to buy'' (Fig.~\ref{fig:phase}(b)).

A simple nonlinearity resembling the form shown in Figure \ref{fig:discriminator} is given by: 
\begin{equation}
\Phi(X)=\frac{\beta X}{1+|\beta X|},\label{eqn:phi}
\end{equation}
and this will be used with $\beta=5$ in the numerical simulations.

\section{Interacting Consumers}

\begin{figure}[h!!]
\centering
\epsfig{file=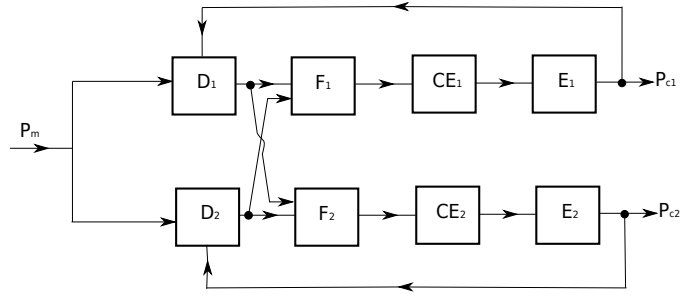,width=0.5\linewidth}
\caption{Model of two coupled consumers.}\label{fig:twocell}
\end{figure}

As consumers interact by exchanging opinions we can consider coupling them through control signals,
the simplest variant being a direct exchange of neighbouring partial MC control signals (Fig.~\ref{fig:twocell}).

To simplify the analysis we assume that for MC$_1$ and MC$_2$, $\Phi_1(X)=\Phi_2(X)\equiv\Phi(X)$, $K_1(p)=K_2(p)\equiv\dfrac{1}{1+ap}$ and $\sigma_1=\sigma_2\equiv\sigma$.
The equations for such coupled consumers can then be written as:
\begin{eqnarray}
\frac{dX_1}{d\tau}+X_1+\Phi(X_1)&=&\gamma_1+\kappa\Phi(X_2),\label{eqn:twocella}\\
\frac{dX_2}{d\tau}+X_2+\Phi(X_2)&=&\gamma_2+\delta\Phi(X_1).\label{eqn:twocellb}
\end{eqnarray}
Here $\tau=\dfrac{t}{a}$ is dimensionless time and $\delta$ and $\kappa$ are the coupling coefficients.
Firstly to decide whether $\delta$ and $\kappa$ should be positive or negative take $\kappa=0$ for simplicity and consider the influence of MC$_1$ on MC$_2$.
When $\gamma_1>0$ we can get the equilibrium state $X^*>0$ from equation (\ref{eqn:twocella}).
This means $\Phi(X_1^*)>0$ and the activity of the first consumer is positive; i.e.: ``to buy''.
Where $\gamma_2>0$, $\delta=0$, the decision ``to buy'' can be obtained from (\ref{eqn:twocellb}), and when $\delta \neq 0$ the opinion of the second consumer needs to be increased.
This means that we need to choose positive signs for the coefficients $\delta$ and $\kappa$.
This can be interpreted as a cooperative type of coupling where the consumers are likely to do the same as their neighbours.
In contrast, negative values for the coupling coefficients would represent an ``antagonistic'' type of interaction, where the neighbours disagree.

\begin{figure}[h!!]
\centering
\epsfig{file=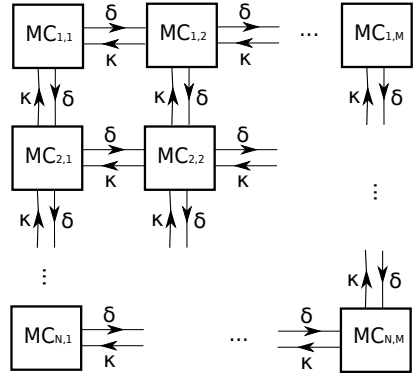,width=0.3\linewidth}
\caption{A 2D lattice network of model consumers (MCs), such as would be seen where people live in a grid road layout.  The 1D chain (considering only a single row of this scheme) could represent a street of consumers.}\label{fig:lattice}
\end{figure}

As social networks, large networks of consumers may have complex structure and emergent complex dynamics. 
To get an insight into the collective Behaviour on networks we focus our attention on regular networks, considering chains and lattices of MCs with nearest neighbour coupling. 
These topologies, in spite of their simplicity, are reasonable models for city lanes and districts, where people living on a 1D or 2D grid-layouts communicate predominantly with their next-door neighbours.
An illustration of the 2D lattice structure is shown in Figure \ref{fig:lattice}.

The other assumption will be that the MCs differ in their initial opinions ($\gamma$) only, while the other characteristics are the same, for example for a 1D chain: $\Phi_n(X)\equiv\Phi(X)$, $K_n(p)\equiv\dfrac{1}{1+ap}$, $\sigma_n\equiv\sigma$, $\kappa_n\equiv\kappa$ and $\delta_n\equiv\delta$, with $n=1,2\ldots N$.

The system of equations for a locally coupled chain can be written as:
\begin{equation}
\frac{dX_n}{d\tau}+X_n+\Phi(X_n)=\gamma_n+\delta\Phi(X_{n-1})+\kappa\Phi(X_{n+1}),\label{eqn:chain}
\end{equation}
where the boundary conditions are $X_0=X_{N+1}=0$.

The dynamical equations for the lattice read:
\begin{eqnarray}
\frac{dX_{n,m}}{d\tau}+X_{n,m}+\Phi(X_{n,m})=\gamma_{n,m}&+&\delta\Phi(X_{n-1,m})+\kappa\Phi(X_{n+1,m})\nonumber\\
&+&\delta\Phi(X_{n,m-1})+\kappa\Phi(X_{n,m+1}),\label{eqn:lattice}
\end{eqnarray}
where $n=1,2\ldots N$, $m=1,2\ldots M$, and the boundary conditions are given by $X_{0,m}=X_{N+1,m}=X_{n,0}=X_{n,M+1}\equiv 0$.

\section{Homogeneous solutions}\label{sec:homog}

Homogeneous solutions arise in (\ref{eqn:chain}) and (\ref{eqn:lattice}) when $\gamma_{n}=\gamma$ or $\gamma_{m,n}=\gamma$ along with $\delta=\kappa$. 
Indeed, under these conditions $X_{n}(t)\equiv X(t)$ and $X_{n,m}(t)\equiv X(t)$ are invariant manifolds of the corresponding dynamical systems. 
Evolution on these manifolds is described by
\begin{equation}
\frac{dX}{d\tau} + X + (1-2d\delta)\Phi(X) = \gamma,\label{eqn:ode_man}
\end{equation}
where $d=1,\ 2$ is the dimensionality of the lattice. 

\begin{figure}[h!!]
\centering
\epsfig{file=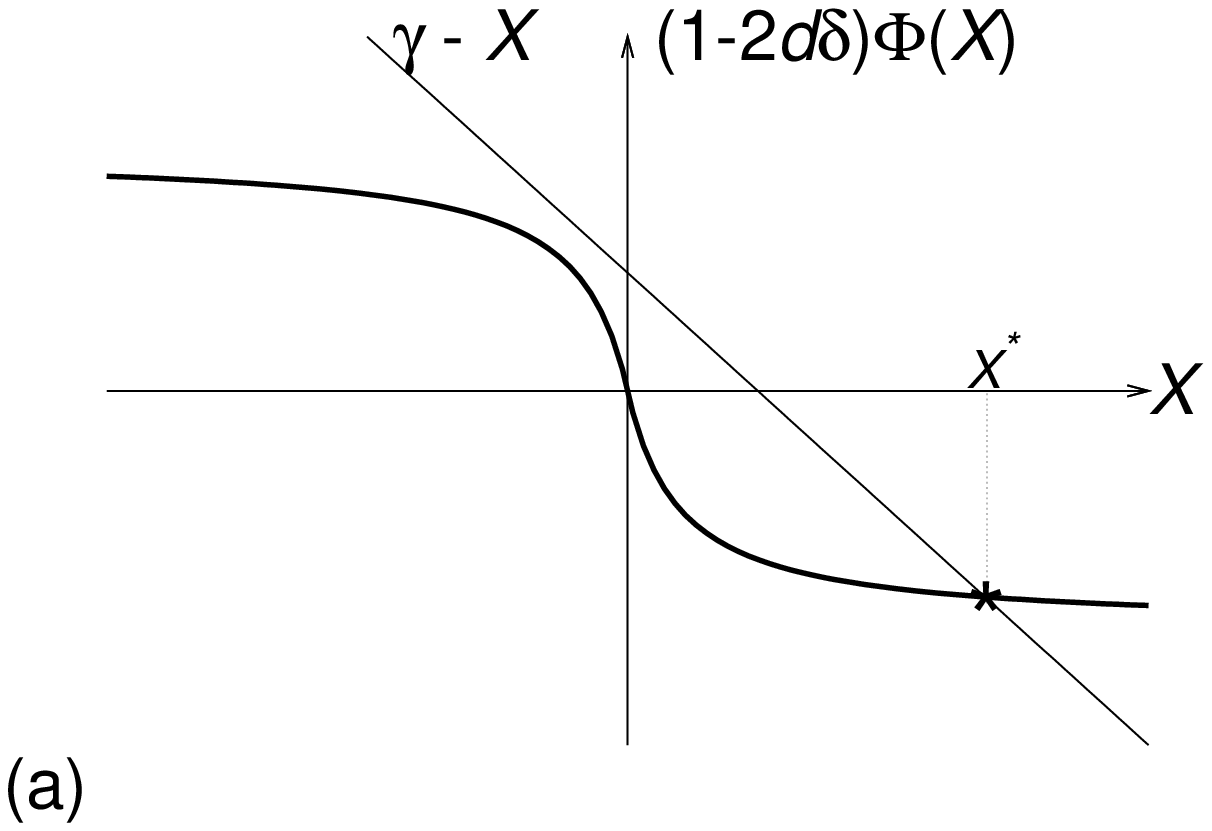,width=0.3\linewidth}
\epsfig{file=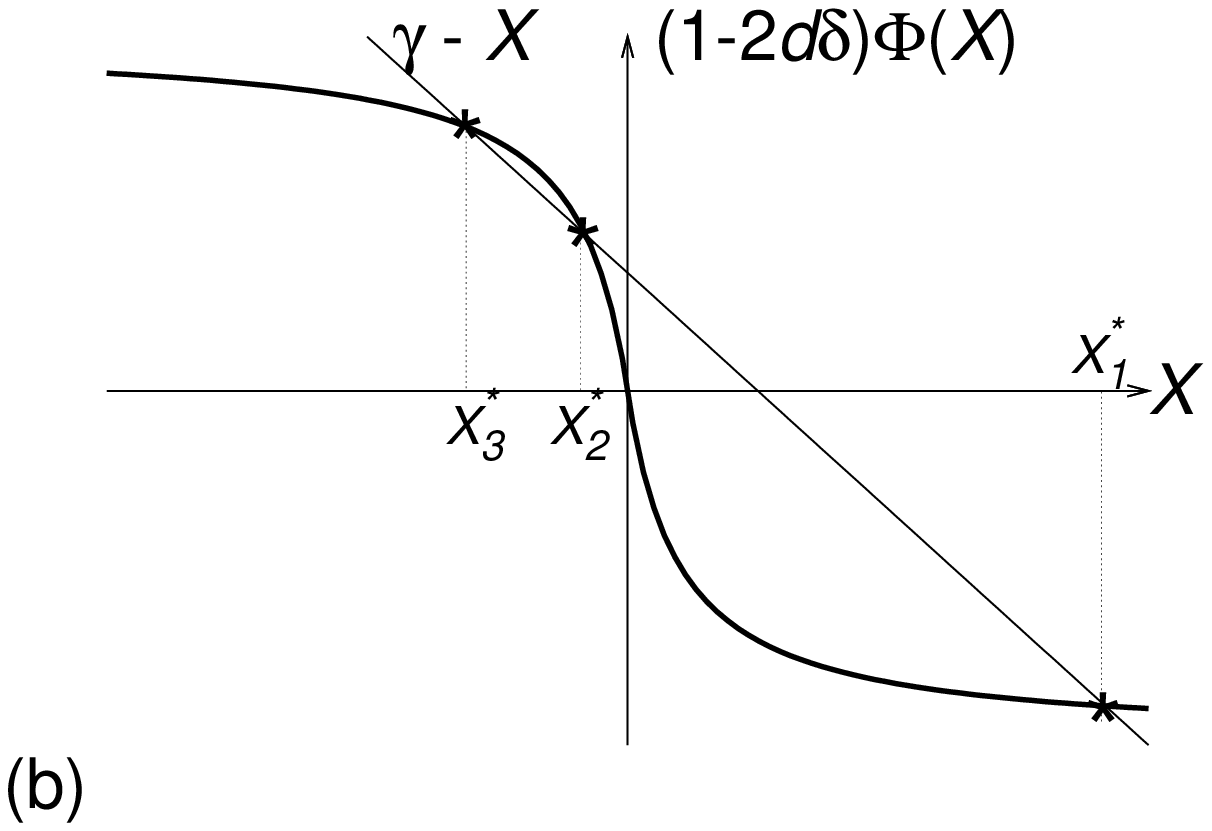,width=0.3\linewidth}
\caption{Possible equilibrium points for model (\ref{eqn:ode_man}) when $2d\delta > 1$.  For different coupling strengths ($\delta=\kappa$) there can exist either one (a) or three (b) equilibria.}\label{fig:phase2}
\end{figure}

First, assume that $\delta>0$. 
If $2d\delta<1$ (\ref{eqn:ode_man}) has a single stable equilibrium as in the case of the autonomous dynamics of a consumer (Fig.~\ref{fig:phase}(a)). 
Notably, in the opposite case there can exist either one (stable) or three (one unstable and two stable) equilibrium points.
Figures \ref{fig:phase2} (a) and (b) illustrate these two cases. 
Approximate solution for the two stable points can be derived is case $|\beta X|\gg 1$, when $\Phi(X)\approx\pm1$. 
Then, 
\begin{equation}\label{eqn:x}
X^*_\pm=\gamma\pm(2d\delta-1),
\end{equation}
 along with the validity condition $|\beta (\gamma\pm(2d\delta-1))|\gg1$. 
The latter can be fulfilled even for small $\gamma$ if the coupling is strong enough.
 
The linear stability of (\ref{eqn:ode_man}) corresponds to the linear stability of (\ref{eqn:chain}) and (\ref{eqn:lattice}) within the invariant manifold. 
This demonstrates that the new points emerge from the existing equilibrium as a result of a pitchfork bifurcation that occurs as the coupling strengthens at some bifurcation value $\delta^*$, the former becoming stable and the latter losing its stability (again, we do not study transversal stability here). 
A precise result is possible for the case of $\gamma=0$, in which 
\begin{equation} \label{eqn:bif}
\delta^*=\frac{1}{2d}\left(1+\frac{1}{\beta}\right).
\end{equation}
Above this threshold the original equations have three spatially homogeneous solutions, among which one is unstable and the others are stable within the invariant manifold. 
Non-zero $\gamma$ changes the basins of attraction of these equilibria in favour of the one with the same sign as $\gamma$. 
It is also easy to see that $\delta^*$ is a monotonically  increasing function of $\gamma$.

Second, if $\delta<0$ one gets 
\begin{equation}
\frac{dX}{d\tau} + X + (1+2d|\delta|)\Phi(X) = \gamma,\label{eqn:ode_man1}
\end{equation}
which always has a single solution. 
However, the ``checkerboard'' manifold $X_{n}(t)=-X_{n-1}(t)=X(t)$ and $X_{n,m}(t)=-X_{n-1,m}(t)=-X_{n+1,m}(t)=-X_{n,m-1}(t)=-X_{n,m+1}(t)=X(t)$ does undergo a pitchfork bifurcation as the dynamics on it obeys
\begin{equation}
\frac{dX}{d\tau} + X + (1-2d|\delta|)\Phi(X) = \gamma. \label{eqn:ode_man2}
\end{equation}
Correspondingly, the bifurcation value for $\gamma=0$ is the same as (\ref{eqn:bif}) except that we now use the absolute value $|\delta|$.

These results suggest the following conjecture on the pattern formation in this system. 
According to (\ref{eqn:x}), strong coupling will result in sharply observed `positive' and `negative' solutions. 
As the initial opinions $\gamma$ are always taken as the initial conditions, one does not observe the bi-stability in numerical experiments, but convergence to one of the stable solutions. 
In the case of the inhomogeneous (random) distributions of $\gamma_n$ or $\gamma_{n,m}$ one can have local prevalence of positive or negative opinions, increasing the `basins of attraction' of the corresponding locally (almost) homogeneous solutions, to which the local dynamics will eventually converge. 
Therefore, one can expect the formation of clusters of positive and negative opinions, possibly, of a complex form in lattices. 
In the case of $\delta<0$ such drastic separation is impossible, as the neighbouring sites exhibit alternation of opinions in any case. 
Still, clusters may be observable, as opinions can have different averages in their neighbourhoods and correspondingly different locally (almost) homogeneous solutions. 

These predictions will be tested in the computational experiments, as shown in section \ref{sec:numerics}.

\section{Localised patterns}\label{sec:patterns}
When small inhomogeneity is added one can expect small perturbations to the homogeneous solution of (\ref{eqn:chain}) and (\ref{eqn:lattice}). 
Furthermore, one can expect that local inhomogeneities will have a local effect under certain conditions. 
The latter will also be a criterion for the absence of large-scale patterns. 

We start by introducing a small perturbation at one site of the chain, while the rest are identical: $\gamma_{n_0}=\gamma+\tilde{\gamma}=\gamma(1+\varepsilon), \ \gamma_n=\gamma \ \forall n\neq n_0$. 
Assuming also that the coupling is symmetric and weak: $\delta=\kappa=\varepsilon \Delta$. 
We now develop  a perturbation theory in powers of small parameter $\varepsilon$, taking the homogeneous stationary solution: 
\begin{equation}
X^* + (1-2\delta)\Phi(X^*) = \gamma,\label{eqn:hom_sol}
\end{equation}
as the zero-order approximation: $X_n=X^*+\varepsilon X_n^{(1)}+\varepsilon^2 X_n^{(2)}+\ldots$. 
We substitute this expansion in the original equations (\ref{eqn:chain}) and find the leading order correction for each site.
In the first order one gets:
\begin{equation}\label{eqn:first_order}
\begin{aligned}
& \varepsilon X^{(1)}_{n_0} + \Phi'(X^*)\varepsilon X_{n_0}^{(1)} = \varepsilon \gamma + \mathcal{O}(\varepsilon^2),\\
& \varepsilon X^{(1)}_{n} + \Phi'(X^*)\varepsilon X_{n}^{(1)}=\mathcal{O}(\varepsilon^2), \ \forall n\neq n_0,
\end{aligned}
\end{equation}
and 
\begin{equation}\label{eqn:first_order_sol}
\begin{aligned}
& X^{(1)}_{n_0} =\frac{ \gamma}{1+\Phi'(X^*)} ,\\
& X^{(1)}_{n} =0, \ \forall n\neq n_0,
\end{aligned}
\end{equation}
In the second order
\begin{equation}\label{eqn:second_order}
\begin{aligned}
& \varepsilon^2 X^{(2)}_{n_0\pm1} + \Phi'(X^*)\varepsilon^2 X_{n_0\pm1}^{(2)} = \varepsilon^2\Delta \Phi'(X^*)X^{(1)}_{n_0} + \mathcal{O}(\varepsilon^3),\\
& \varepsilon^2 X^{(2)}_{n} + \Phi'(X^*)\varepsilon^2 X_{n}^{(1)}=\mathcal{O}(\varepsilon^3), \ \forall n\neq n_0, n_0\pm1,
\end{aligned}
\end{equation}
and 
\begin{equation}\label{eqn:second_order_sol}
\begin{aligned}
& X^{(2)}_{n_0} =\frac{\Delta \Phi'(X^*)}{1+\Phi'(X^*)} X_{n_0}^{(1)}=\frac{\Delta \Phi'(X^*)}{1+\Phi'(X^*)} \frac{ \gamma}{1+\Phi'(X^*)} ,\\
& X^{(2)}_{n} =0, \ \forall n\neq n_0, n_0\pm1,
\end{aligned}
\end{equation}

Finally, one obtains the leading order perturbations:
\begin{equation}\label{eqn:decay}
X_{n_0\pm l}-X^*=\lambda^l \frac{ \tilde\gamma}{1+\Phi'(X^*)}+\mathcal{O}(\varepsilon^{l+2}), \ \lambda=\frac{\delta \Phi'(X^*)}{1+\Phi'(X^*)},
\end{equation}
The perturbations are exponentially Localised if $\lambda<1$ and have the localisation length $L=-\frac{1}{\ln{\lambda}}, \ X_{n_0\pm l}-X^*=(X_{n_0}-X^*)e^{-l/L}$. 
In this case we predict the stationary pattern of small-scale fluctuations in the general case of random $\gamma_n$. 
In the opposite case even a small local perturbation will cause the same order perturbations over the whole system and the stationary patterns may get coarse-grained, their size potentially becoming of the order of the system size. 
By the estimate $\lambda=\frac{\delta \Phi'(X^*)}{1+\Phi'(X^*)}\le \frac{\delta \beta}{1+\beta}$ one can conclude that the generic route to pattern coarse-graining is through coupling strengthening, the critical value being $\delta^{**}=1+\frac{1}{\beta}$. 

Remarkably, the bistability of the homogeneous solution threshold (\ref{eqn:bif}) is of the same order.
From this we predict that large scale clusters of sharply positive and negative opinions will form if the coupling is stronger than $\delta^*, \ \delta^{**}$.

It is straightforward to show that the same expression for the decay (\ref{eqn:decay}) is valid for the horizontal and vertical directions in lattices. Thus, the same localisation criterion applies and different pattern formation regimes will take place in the similar regions of the parameter space.

The case $\beta<0$ and perturbations to the checkerboard solution can be analysed similarly and the same criteria can be derived.

\section{Numerical results}\label{sec:numerics}
Computer simulations were carried out to investigate the Behaviour of chains and lattices of consumers by the numerical integration of  (\ref{eqn:chain}) and (\ref{eqn:lattice}).
The results described here for both cases are presented in Figures \ref{fig:results1}--\ref{fig:results3}.

\subsection*{1D Chains}

In all cases for the chain the number of consumers in the simulation of the model was chosen to be $N=20$ for ease of visualization.
Different initial conditions (distributions of $\gamma_n$) were used for each of the figures and are shown in sub-figure (a) in each case.
Sub-figures (c1)--(f1) show the long-time (post-transient) results of having different coupling interactions; in each case $\delta,\kappa=0.3, 2, -0.8, -2$, respectively.

\begin{figure}[f]
\epsfig{file=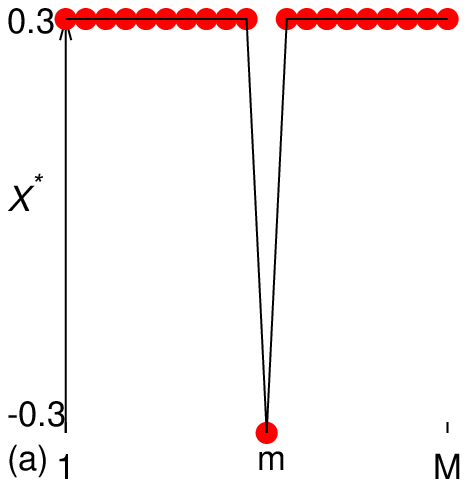,width=0.18\linewidth}
\epsfig{file=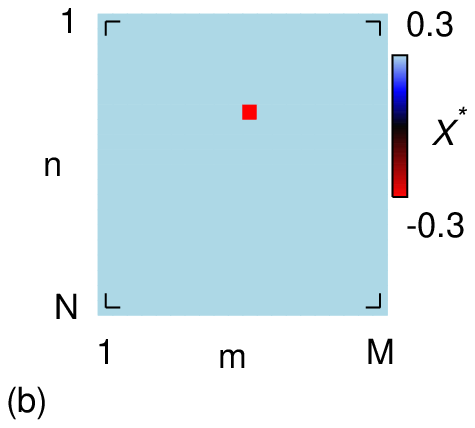,width=0.2\linewidth}

\epsfig{file=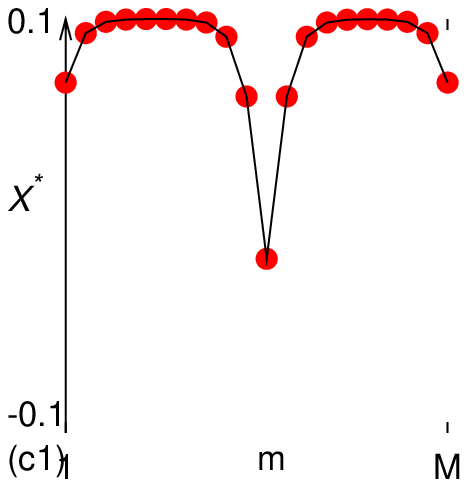,width=0.18\linewidth}
\epsfig{file=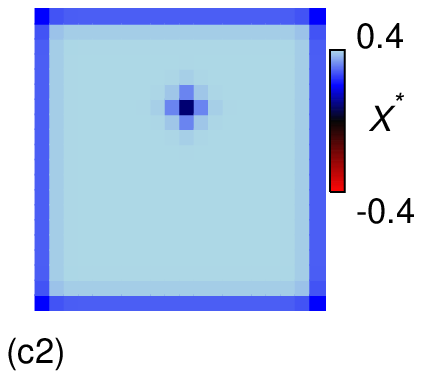,width=0.2\linewidth}
\epsfig{file=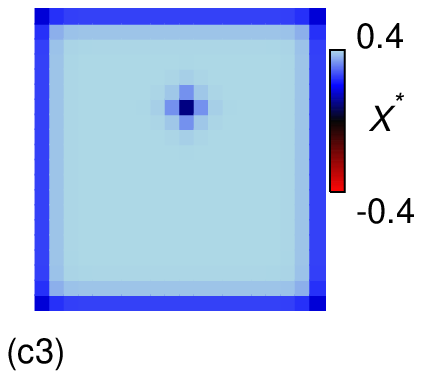,width=0.2\linewidth}
\epsfig{file=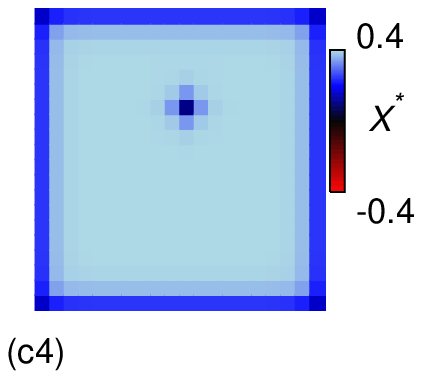,width=0.2\linewidth}
\epsfig{file=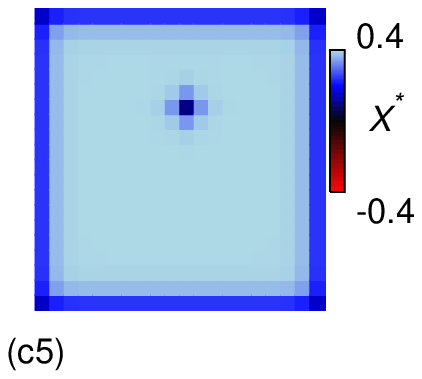,width=0.2\linewidth}

\epsfig{file=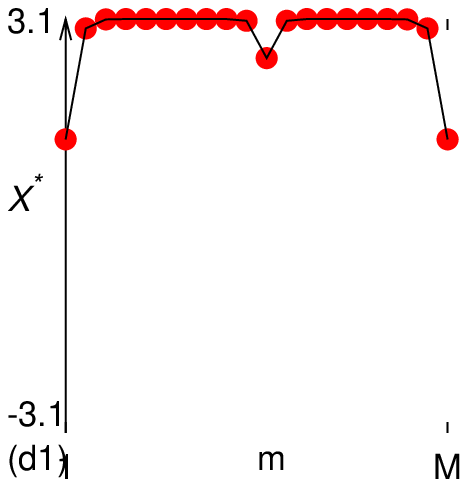,width=0.18\linewidth}
\epsfig{file=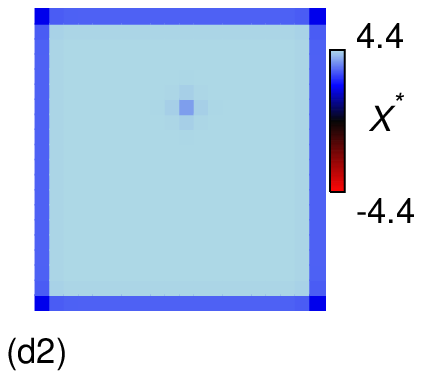,width=0.2\linewidth}
\epsfig{file=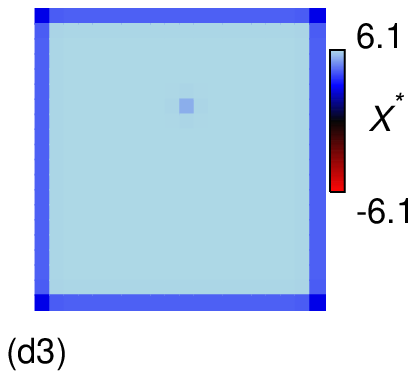,width=0.2\linewidth}
\epsfig{file=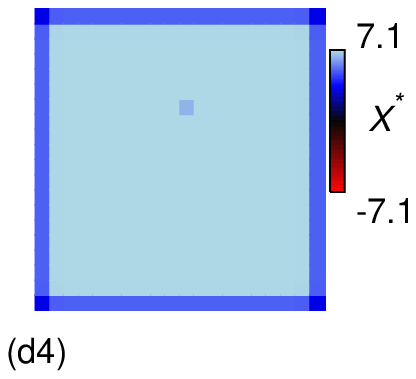,width=0.2\linewidth}
\epsfig{file=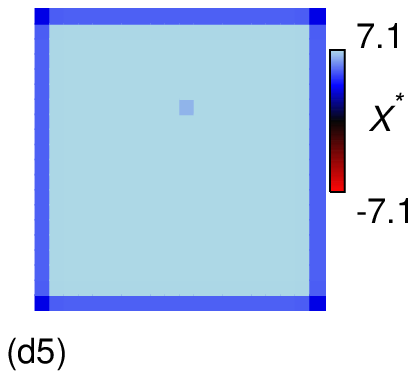,width=0.2\linewidth}

\epsfig{file=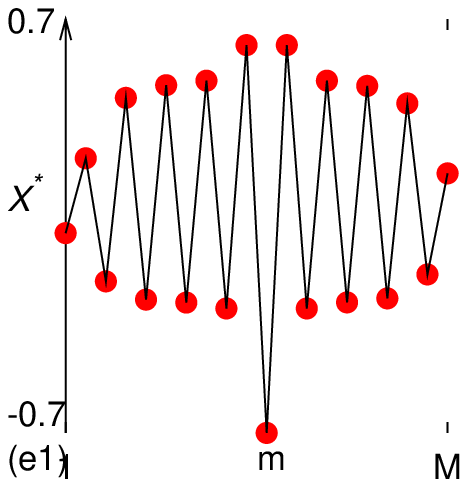,width=0.18\linewidth}
\epsfig{file=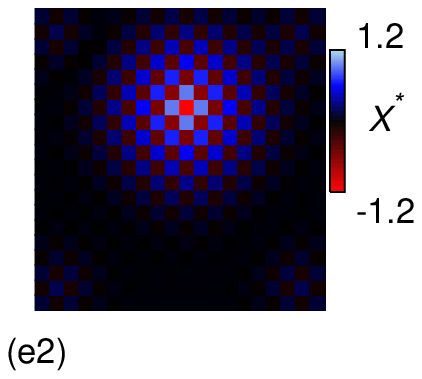,width=0.2\linewidth}
\epsfig{file=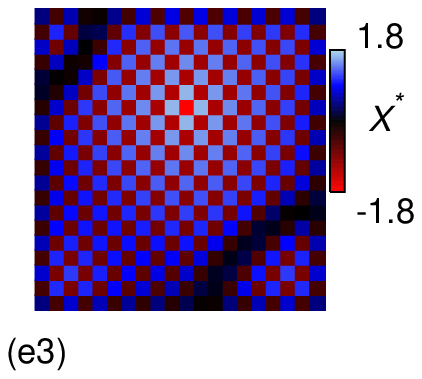,width=0.2\linewidth}
\epsfig{file=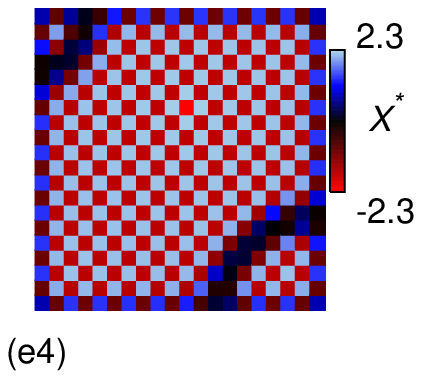,width=0.2\linewidth}
\epsfig{file=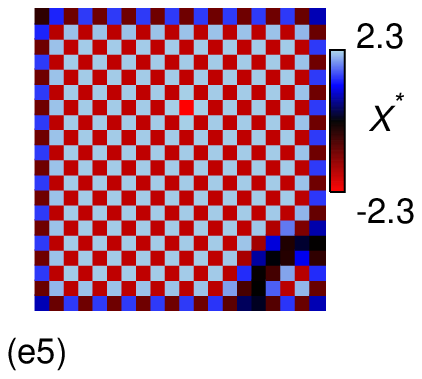,width=0.2\linewidth}

\epsfig{file=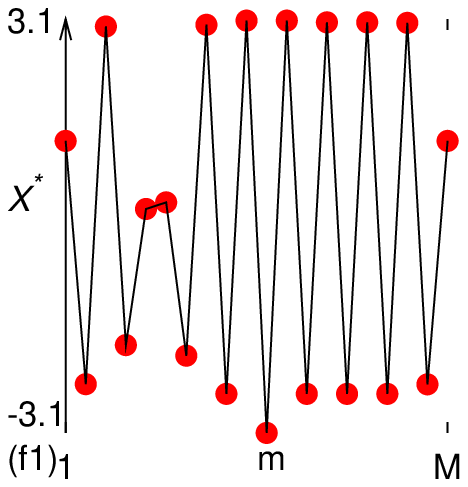,width=0.18\linewidth}
\epsfig{file=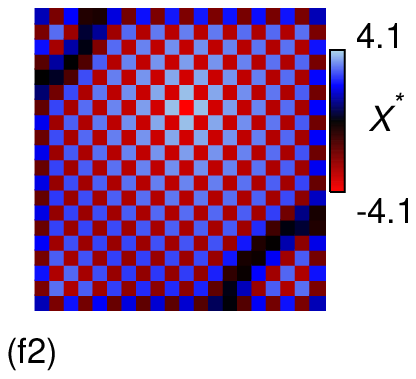,width=0.2\linewidth}
\epsfig{file=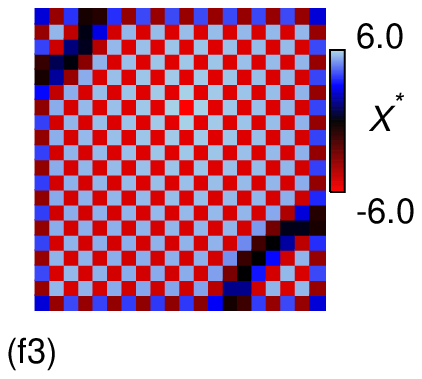,width=0.2\linewidth}
\epsfig{file=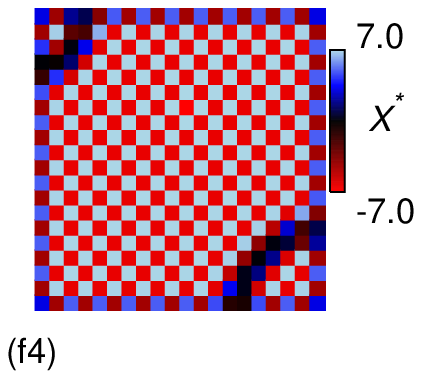,width=0.2\linewidth}
\epsfig{file=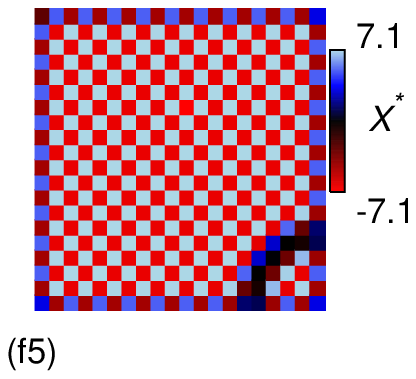,width=0.2\linewidth}

\caption{The initial condition, here a single perturbed site in an otherwise uniform field, is shown for both the chain (a) and lattice (b). In cases (c), (d), (e), (f)  the coupling strength $\delta=\kappa=0.3, 2, -0.8, -2$, respectively.  Column 1 shows the post-transient state of the chain at $\tau =20$. The evolution of the state of the lattice is shown at $\tau =1,2,5,20$ in columns 2, 3, 4 and 5, respectively. Weak cooperative coupling ($\delta=\kappa=0.3$) slightly smooths the variation across neighbours but they retain their basic distribution.  Stronger coupling ($\delta=\kappa=2$) results in a more uniform final distribution, with the original bias of the perturbed individual only just discernible.  For antagonistic coupling a checkerboard pattern emerges, with neighbouring individuals in opposition.  In the weaker coupling case $\delta=\kappa=-0.8$ the effect of the perturbed individual can be seen more clearly than in the stronger case ($\delta=\kappa=-2$) which results in a more uniform pattern.}\label{fig:results1}
\end{figure}

For the results in Figure \ref{fig:results1} all $\gamma_n$ were chosen to be 0.3 other than a single perturbed site at $n=10$, taken as $\gamma_{10}=-0.3$.
This can be considered as investigating the effect of having a single consumer with a deviant opinion in the system.
It can be seen that weak cooperative coupling slightly smooths the variation across neighbours over time.
However, they can be seen to retain their basic distribution; a result of holding on to their original opinion (as $\gamma_n(t)=const$).  
Stronger cooperative coupling results in a more uniform final distribution, with a small effect of the perturbed individual.  
Unsurprisingly, for antagonistic coupling a pattern of opposing neighbours is seen to emerge as predicted in Section \ref{sec:patterns}.  
In the weaker coupling case the effect of the perturbation can be seen more clearly than in the stronger case; with the latter resulting in a more uniform pattern.
As will be seen in Section \ref{sec:localisation}, there is a coupling value, above which the uniform pattern emerges but below which shows localisation of the pattern; again in accordance with the results in Section \ref{sec:patterns}.
This is related to the qualitative change in Behaviour described next.

\begin{figure}[f]
\epsfig{file=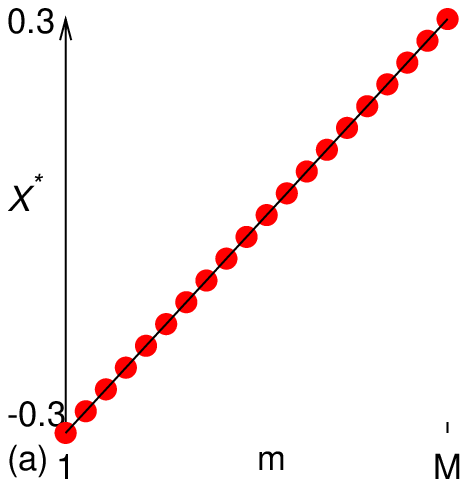,width=0.18\linewidth}
\epsfig{file=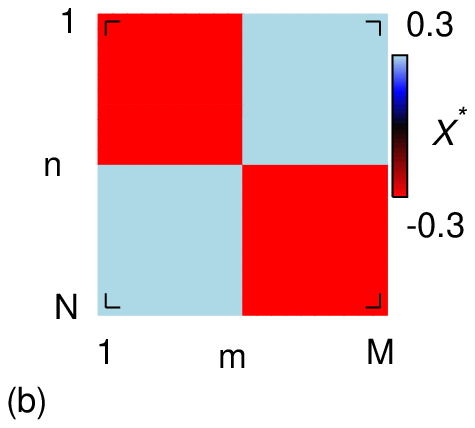,width=0.2\linewidth}

\epsfig{file=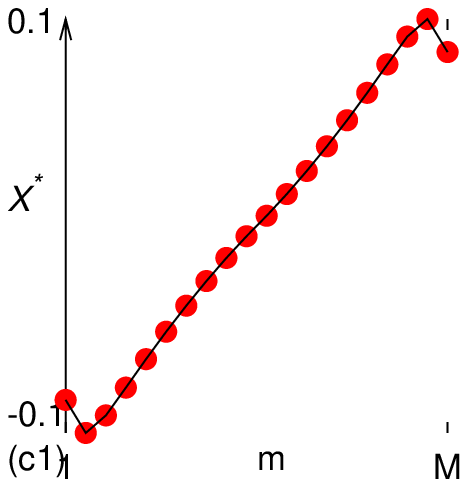,width=0.18\linewidth}
\epsfig{file=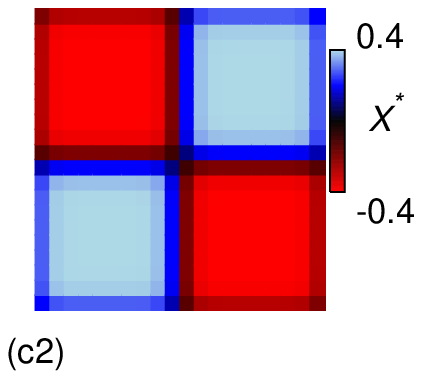,width=0.2\linewidth}
\epsfig{file=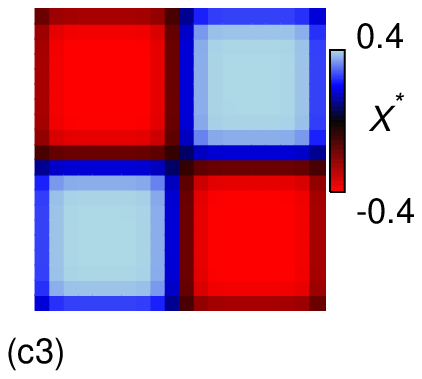,width=0.2\linewidth}
\epsfig{file=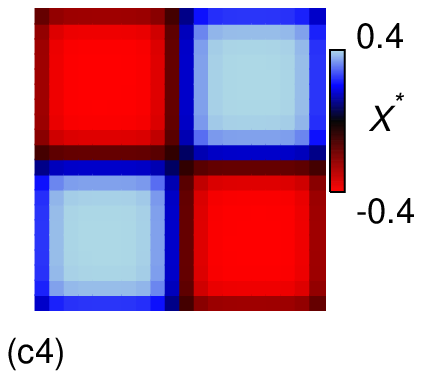,width=0.2\linewidth}
\epsfig{file=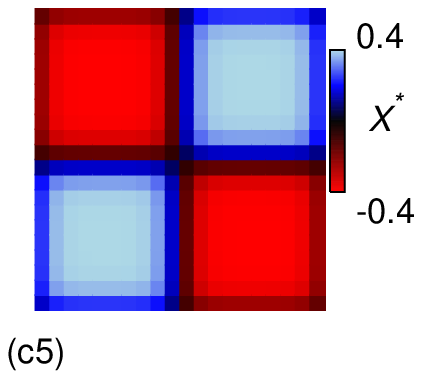,width=0.2\linewidth}

\epsfig{file=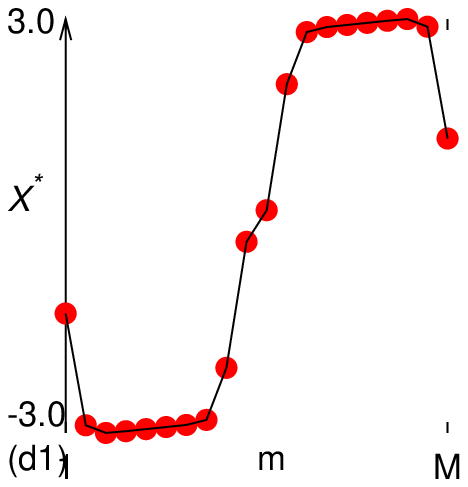,width=0.18\linewidth}
\epsfig{file=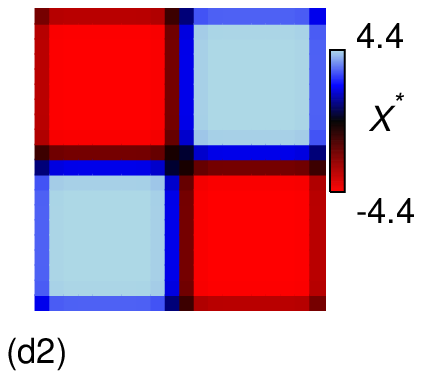,width=0.2\linewidth}
\epsfig{file=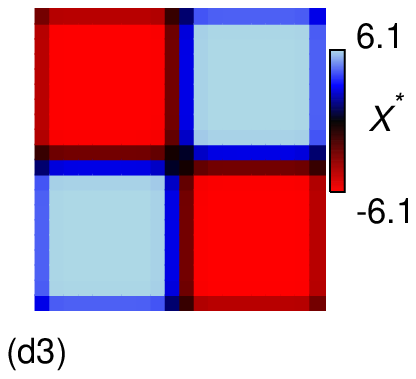,width=0.2\linewidth}
\epsfig{file=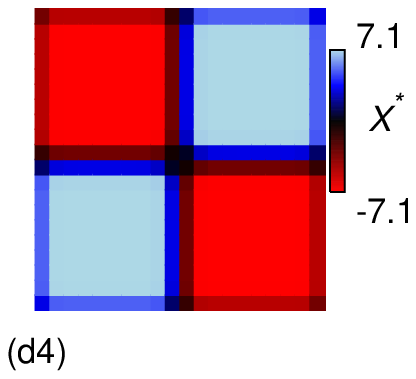,width=0.2\linewidth}
\epsfig{file=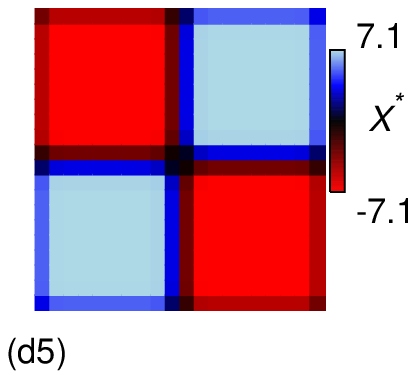,width=0.2\linewidth}

\epsfig{file=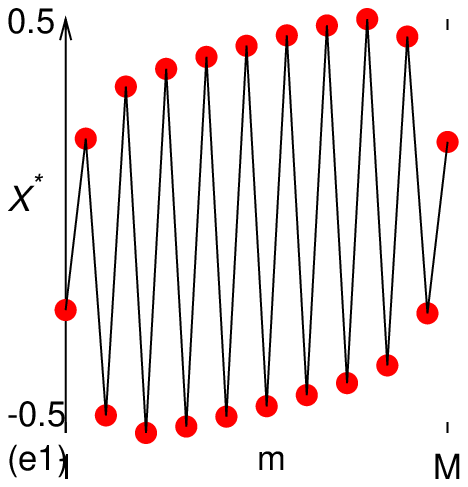,width=0.18\linewidth}
\epsfig{file=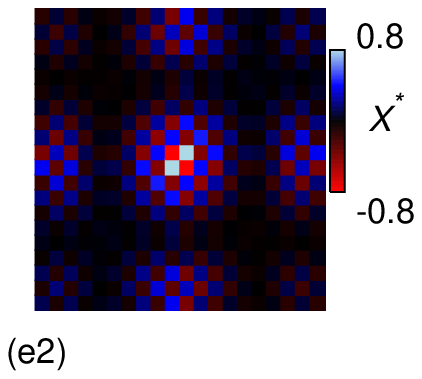,width=0.2\linewidth}
\epsfig{file=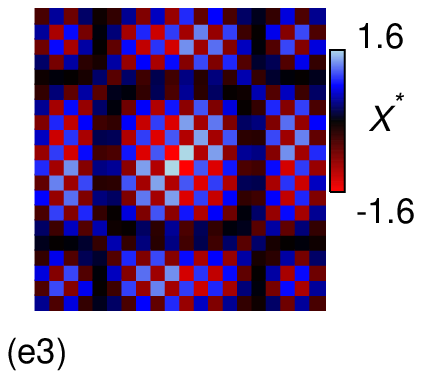,width=0.2\linewidth}
\epsfig{file=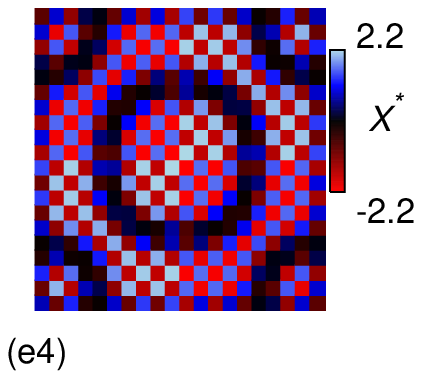,width=0.2\linewidth}
\epsfig{file=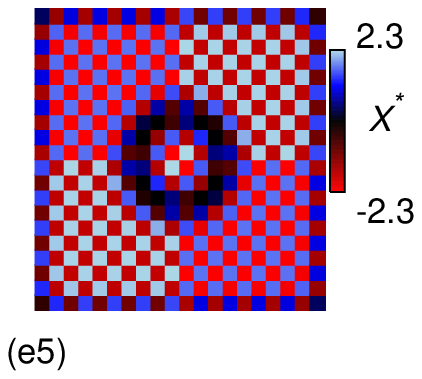,width=0.2\linewidth}

\epsfig{file=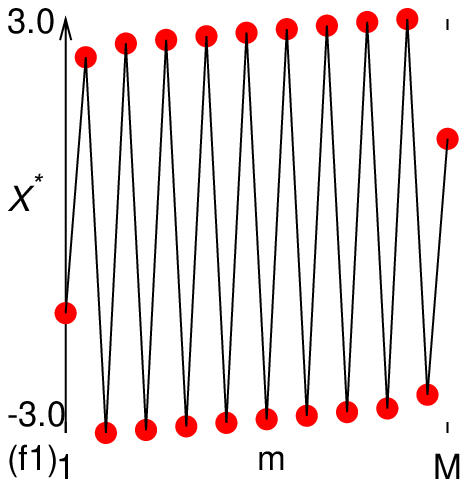,width=0.18\linewidth}
\epsfig{file=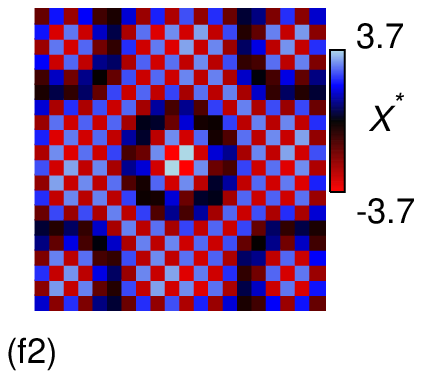,width=0.2\linewidth}
\epsfig{file=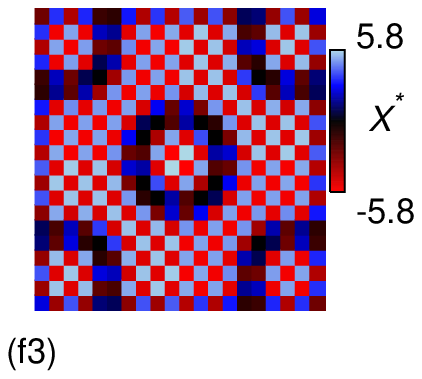,width=0.2\linewidth}
\epsfig{file=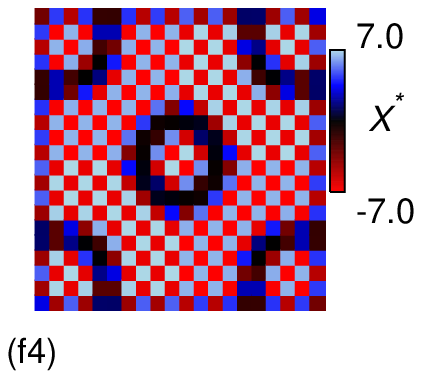,width=0.2\linewidth}
\epsfig{file=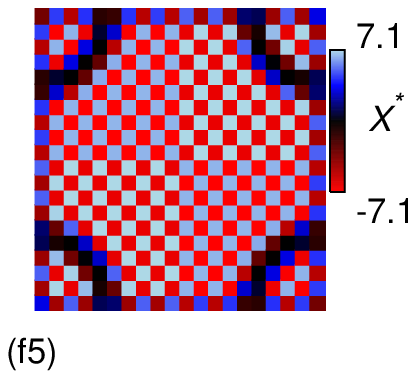,width=0.2\linewidth}

\caption{The initial distribution of a linear variation of opinions between $-0.3\leq\gamma\leq0.3$ for a chain (a), and a coarse (10 cell block) checkerboard pattern for the lattice (b). In cases (c), (d), (e), (f)  the coupling strength $\delta=\kappa=0.3, 2, -0.8, -2$, respectively.  Column 1 shows the post-transient state of the chain at $\tau =20$. The lattice is shown at $\tau =1,2,5,20$ in columns 2, 3, 4 and 5, respectively. Weak cooperative coupling results in distributions similar to the initial states, but with different magnitude.  For strong cooperative coupling individuals end up in one of two clusters, $X^*\approx-3$ or $X^*\approx3$, originating from $\gamma_n<0$ or $\gamma_n>0$, respectively. For negative coupling the results are much as in the previous case, with strong coupling showing a more uniform pattern than the weak coupling, which shows the original opinion more clearly. }\label{fig:results2}
\end{figure}

The second initial distribution considered for the chain is a linear variation of opinions between $-0.3\leq\gamma\leq0.3$ (Fig.~\ref{fig:results2} (a)).
This is a situation where near-neighbours have similar opinions but individuals at a distance from one-another originally differ in their estimate of the fair price.
For antagonistic coupling the results are much as in the previous case, with strong coupling showing a more uniform `checkerboard' pattern and weak coupling revealing the underlying bias (distribution of $\gamma_n$) more clearly.
However, here the difference between the final distributions are more interesting for cooperative coupling at the two different strengths.
Where individuals only weakly interact the form of the distribution remains largely unchanged from their original estimate, with only a change in the magnitude and slope, with a roughly linear variation across the range $-0.1\leq X^* \leq0.1$.
In the case of individuals being strongly influenced by their neighbours a different pattern emerges, with consumers end up divided into two clusters of approximately equal opinion as conjectured from Equation (\ref{eqn:x}) in Section \ref{sec:homog}.
These clusters are divided sharply at the interface, with those whose underlying prejudice was negative in one ($X^*\approx-3$) group and those with an originally positive estimate in the other ($X^*\approx3$).
This is in accordance with the predictions made at the end of Section \ref{sec:homog}.
Clearly a qualitative change in the decision-making Behaviour occurs above a certain coupling strength, as will be seen more dramatically in the case of 2D lattices and studied in more detail in Sections \ref{sec:localisation} and \ref{sec:clusters}.

\begin{figure}[f]
\epsfig{file=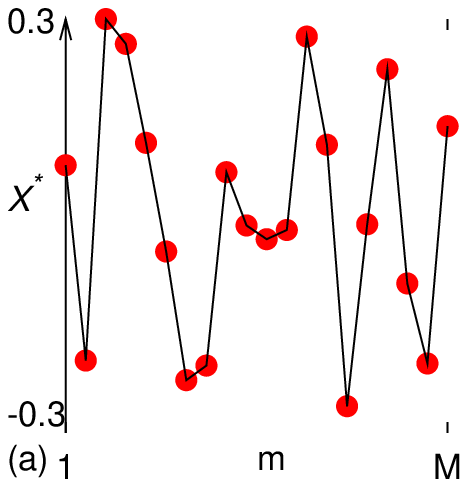,width=0.18\linewidth}
\epsfig{file=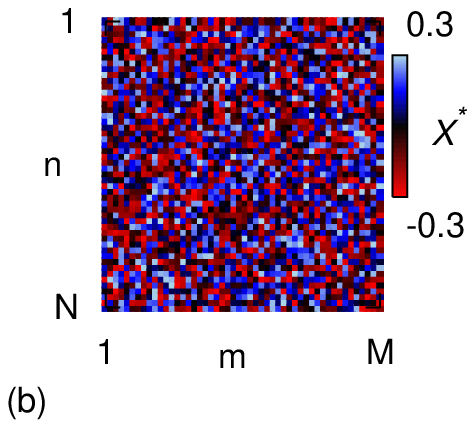,width=0.2\linewidth}

\epsfig{file=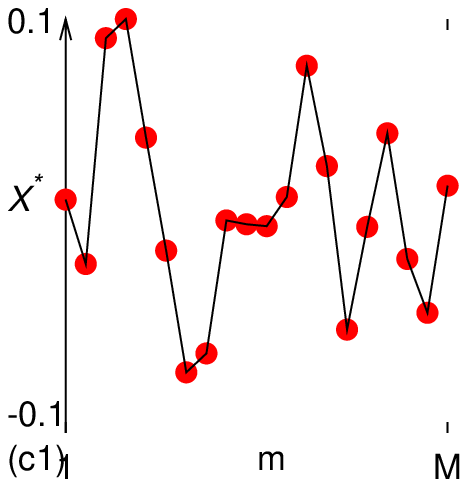,width=0.18\linewidth}
\epsfig{file=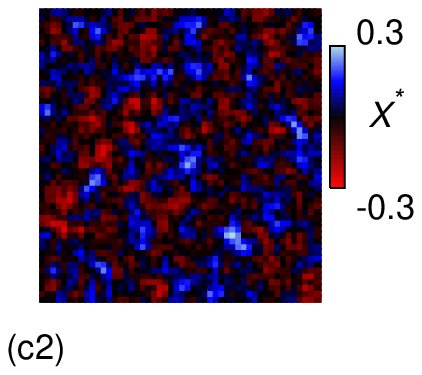,width=0.2\linewidth}
\epsfig{file=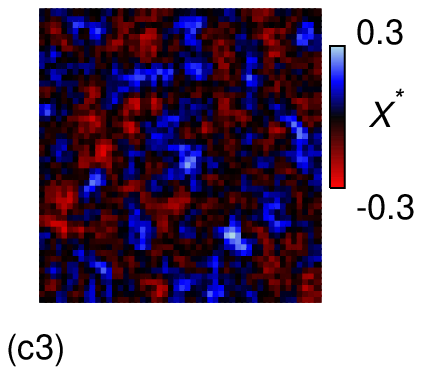,width=0.2\linewidth}
\epsfig{file=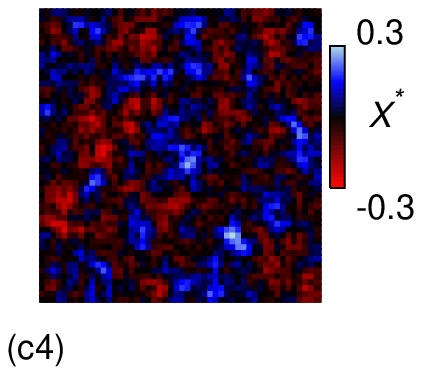,width=0.2\linewidth}
\epsfig{file=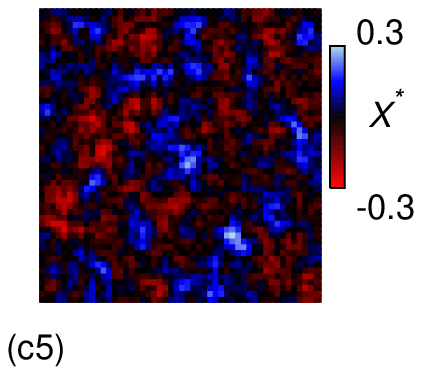,width=0.2\linewidth}

\epsfig{file=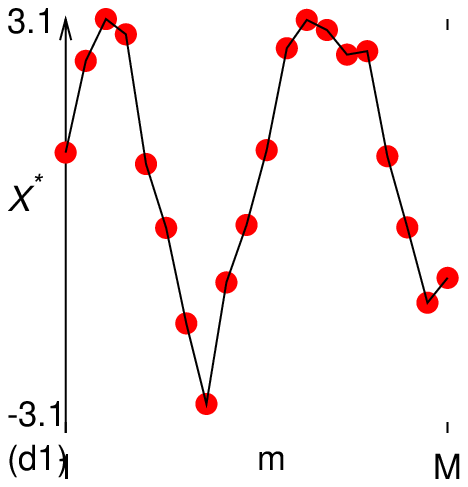,width=0.18\linewidth}
\epsfig{file=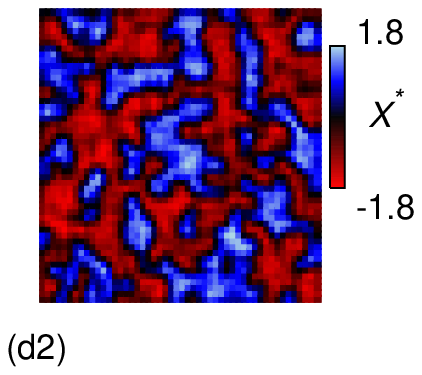,width=0.2\linewidth}
\epsfig{file=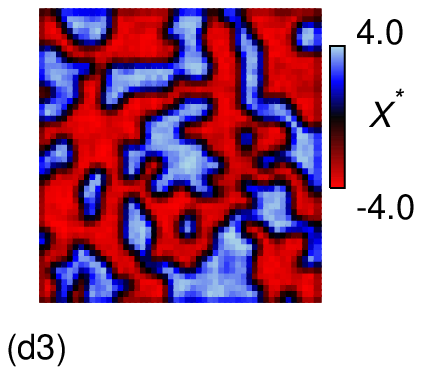,width=0.2\linewidth}
\epsfig{file=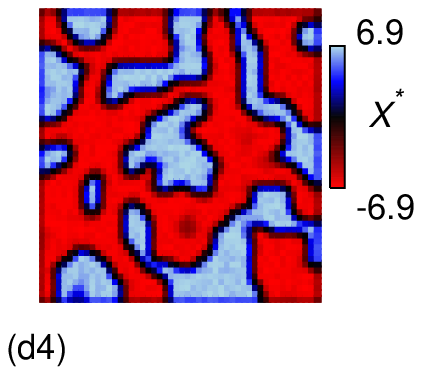,width=0.2\linewidth}
\epsfig{file=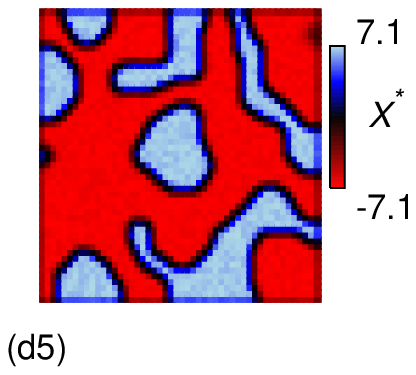,width=0.2\linewidth}

\epsfig{file=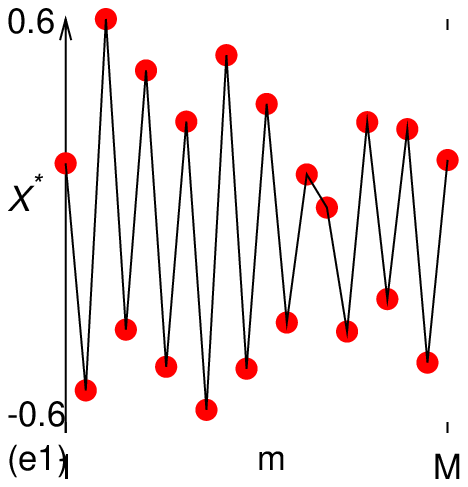,width=0.18\linewidth}
\epsfig{file=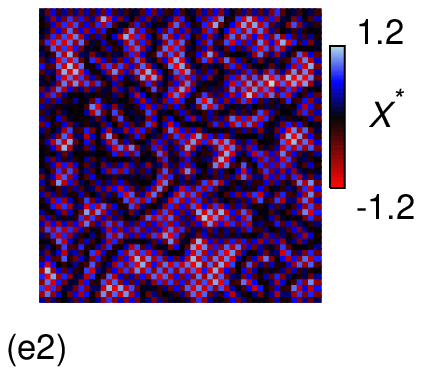,width=0.2\linewidth}
\epsfig{file=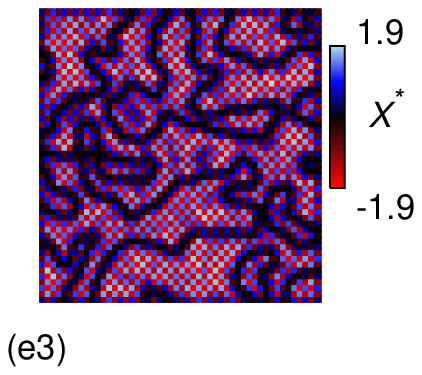,width=0.2\linewidth}
\epsfig{file=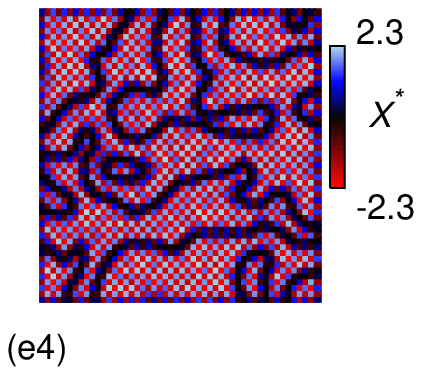,width=0.2\linewidth}
\epsfig{file=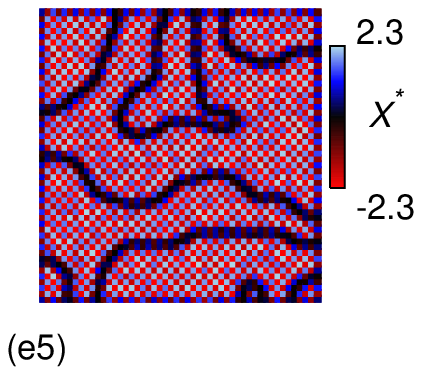,width=0.2\linewidth}

\epsfig{file=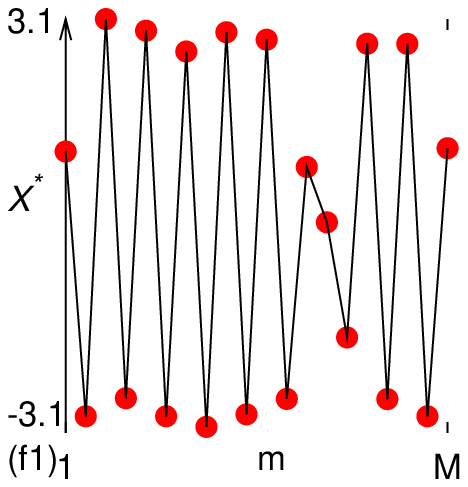,width=0.18\linewidth}
\epsfig{file=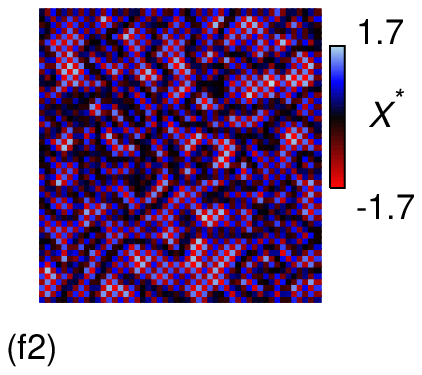,width=0.2\linewidth}
\epsfig{file=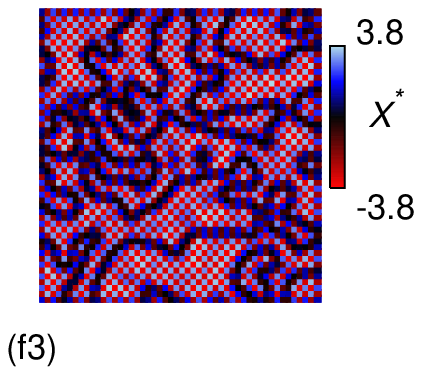,width=0.2\linewidth}
\epsfig{file=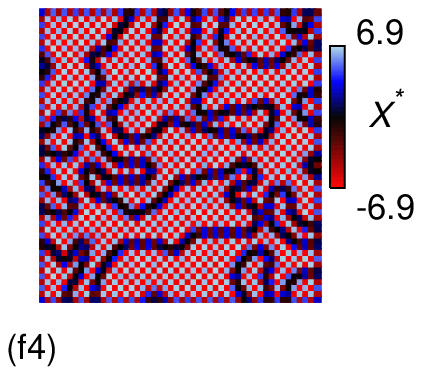,width=0.2\linewidth}
\epsfig{file=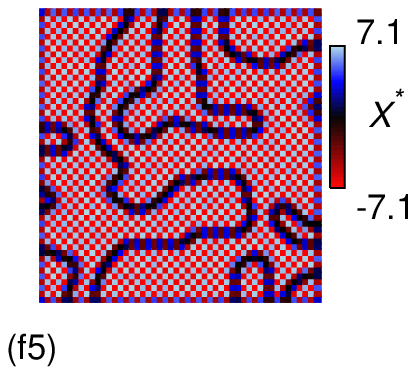,width=0.2\linewidth}

\caption{The initial condition of a random field for the chain (a) and lattice (b). As before (c), (d), (e), (f)  show the coupling strength $\delta=\kappa=0.3, 2, -0.8, -2$, respectively.  Also column 1 shows the state of the chain at $\tau =20$ and the state of the lattice is shown at $\tau =1,2,5,20$ in columns 2, 3, 4 and 5, respectively. In this case a $50\times50$ grid was used for better visualization. Clear domains of clusters are seen to emerge. For negative coupling the clusters are separated by domain boundaries defined by lines of frustration. In the case of weak cooperative coupling ($\delta=\kappa=0.3$) the formation of weak clusters resembling the original, which do not go on to merge together, are formed. For strong coupling ($\delta=\kappa=2$) the clusters are much more well defined and display clear coarsening Behaviour in the lattice.}\label{fig:results3}
\end{figure}

A more natural starting condition, before the neighbours start to exchange information on their opinions, would be an initially arbitrary distribution.
This is considered for the results shown in Figure \ref{fig:results3}, where the distribution of $\gamma_n$ was obtained using a pseudo-random number generator.
Again the antagonistic coupling shows similar Behaviour to above, and qualitatively similar Behaviour to the previous case can be seen for cooperative coupling.
In this case the distribution retains its random nature from the initial distribution, as shown theoretically in Section \ref{sec:patterns}.
However, in the strong coupling case the tendency towards clusters can be clearly seen; predicted as locally homogeneous solutions in Section \ref{sec:homog} and analysed in Section \ref{sec:patterns}.
This clustering Behaviour appears in a more spectacularly in the case of two dimensional lattices, shown next.

\subsection*{2D Lattices}

Numerical results for lattices were carried out using the same discriminator function (\ref{eqn:phi}) as before (also shown in figures \ref{fig:results1}--\ref{fig:results3}).
The initial conditions for the lattice, shown in sub-figure (b) in each case, are: for Figure \ref{fig:results1}, a single perturbed site in an otherwise uniform field; Figure \ref{fig:results2}, a tiling dividing the field into four alternating regions of negative and positive opinions; and Figure \ref{fig:results3}, a pseudo-random field of $\gamma_{n,m}$.

In the first case (Fig.~\ref{fig:results1}) the results are completely analogous to the results for a chain seen above and predicted in Section \ref{sec:patterns}, with smoother final distributions for stronger coupling and antagonistic coupling leading to a checkerboard of opposing neighbours.
There is evidence of frustration at the corners in the latter case, as also seen in Figure \ref{fig:results1} (f1) for sites 5--6 in a chain of antagonistic consumers.
The larger region is seen to dominate to push the system towards a single `cluster' at later times (in the top left corner), similar to the coarsening Behaviour described below.
The second case here (fig.~\ref{fig:results2}) is much less noteworthy; particularly for the cooperative case, given that the initial condition resembles the final state very closely in form, varying only in magnitude.

However, for the irregular initial distributions the results shown in Figure \ref{fig:results3} are far more interesting.
In this case a larger system of $N,M=50$ was used to display the effect more strikingly.
Here, clear domains of clusters are seen to emerge for stronger coupling strengths, as seen in the previous case and in agreement with the previous results.
In the antagonistic case the clusters are the same as each other (opposing neighbours) but separated by domain boundaries defined by lines of frustration.
Coarsening of the patterns can be seen more strikingly in the case of cooperative coupling, in accordance with the predictions made in Sections \ref{sec:homog} and \ref{sec:patterns}.
Weak coupling ($\delta,\kappa=0.3$) results in the formation of small irregular clusters, resembling the original random distributions, which do not go on to merge together over time.
In the case of stronger coupling ($\delta,\kappa=2$) the clusters are much more well defined, with local groups of individuals taking strongly negative or positive opinions and having sharp domain-boundaries, as predicted in Section \ref{sec:patterns}.
For these values the system also displays clear coarsening Behaviour, with groups of similar opinion growing in size until one dominates.
The clustering and coarsening Behaviour is investigated in more detail along with quantification of the phenomena in section \ref{sec:clusters}.

\section{Transition Between Localised and Global Patterns of Behaviour}

\subsection{localisation and De-localisation in the 1D Chains}\label{sec:localisation}

\begin{figure}
\centering
\epsfig{file=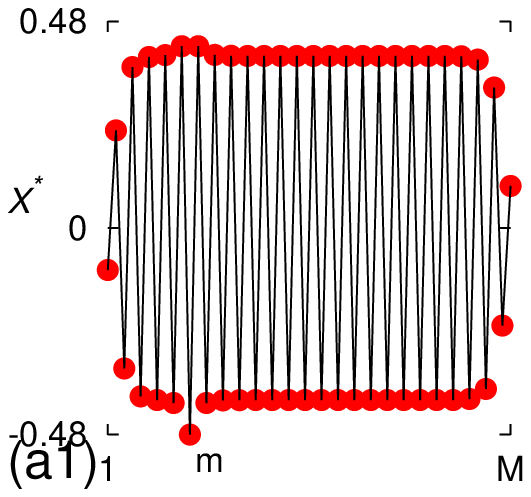,width=0.24\linewidth}
\epsfig{file=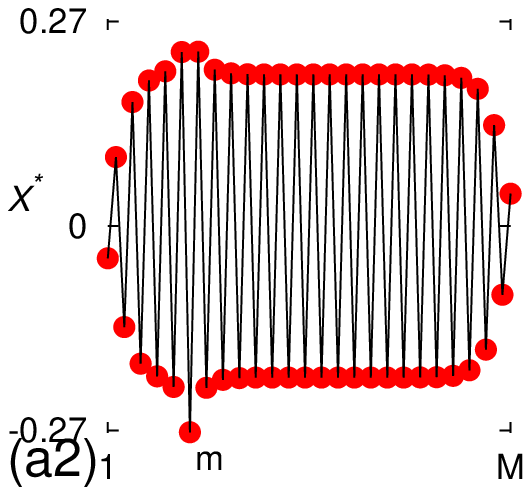,width=0.24\linewidth}
\epsfig{file=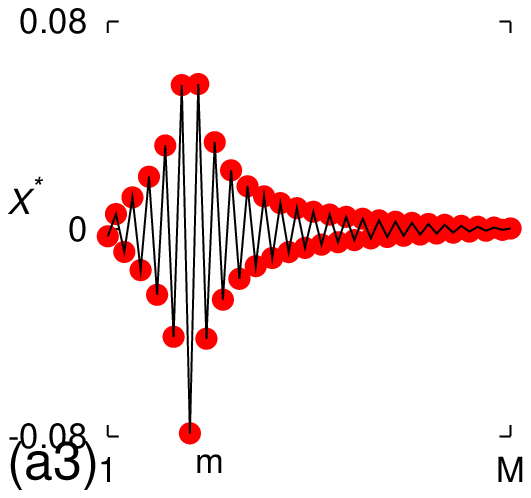,width=0.24\linewidth}
\epsfig{file=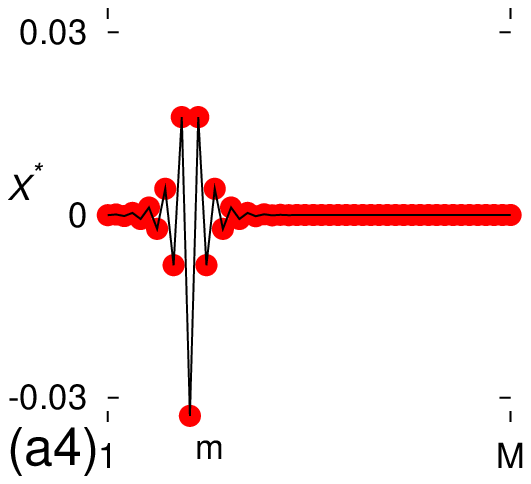,width=0.24\linewidth}

\epsfig{file=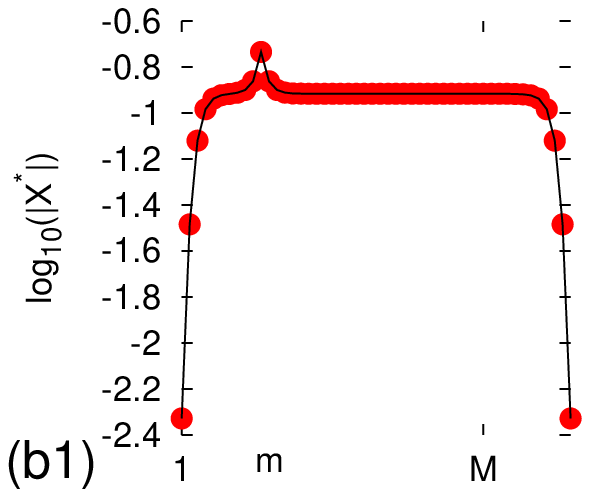,width=0.24\linewidth}
\epsfig{file=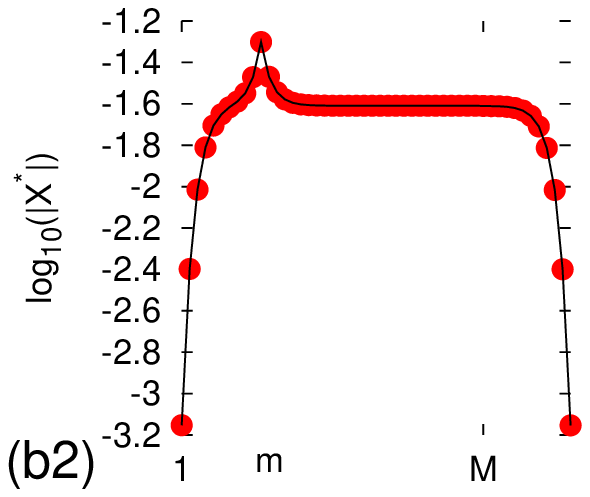,width=0.24\linewidth}
\epsfig{file=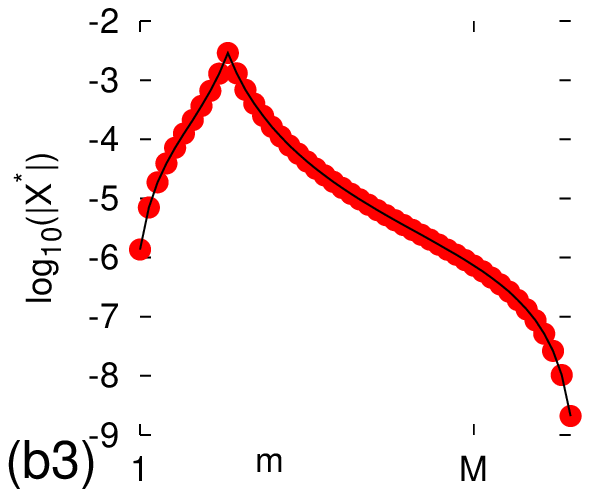,width=0.24\linewidth}
\epsfig{file=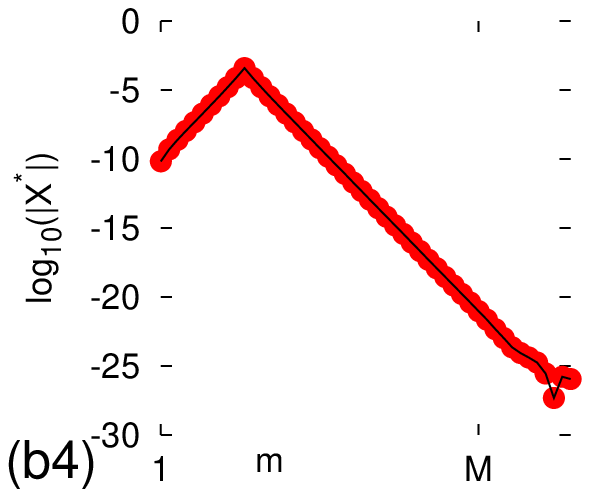,width=0.24\linewidth}

\caption{localisation of the perturbation for different coupling values, on a linear scale (a) and a logarithmic scale (b).  The coupling coefficients are $\delta=\kappa=-0.8, -0.7, -0.6, -0.5$ for columns 1--4, respectively.  A transition between de-Localised and Localised Behaviour occurs.}\label{fig:loc}
\end{figure}

Here we attempt to quantify the localisation of the perturbation as seen in the previous results.
This transition is very important, given the underlying intent of the model to support initiatives on energy sustainability by local authorities, whose decisions would be based on the response of the group rather than the individual consumers.
This quantification is carried out on the one dimensional chain, looking at the degree of localisation on variation of the coupling strength.
Chains containing $N=40$ individuals were used with a single site perturbed over a range of perturbation size $\gamma_j$, where all other $\gamma_n=0$.
Figure \ref{fig:loc} shows the effect for $\gamma_j=0.1$ at coupling strengths $\delta,\kappa=-0.8, -0.7, -0.6, -0.5$, on both a linear and a logarithmic scale.
Qualitatively different distributions can be seen either side of a transition, one with a Localised disturbance and the other de-Localised.
For small coupling parameter the decay is exponential but otherwise it is more irregular and falls to a steady state.  
It can be seen that, for very small coupling, the localisation length could be well defined by a localisation length $L$ in the expression $X_n=\exp(\frac{i}{L})$.

As a measure of localisation we employ the `participation number' $P$.
This is obtained by calculating the normalized stationary amplitudes of $X$:
$$z_n=\frac{|X_n|}{\sum_n |X_n|},$$ 
where $|z_n|$ sums to unity. 
The participation number then is given by the formula:
$$P=\frac{1}{\sum z_n^2}.$$
This is a measure of the effective number of sites that get ``substantially'' perturbed. 

\begin{figure}
\centering
\epsfig{file=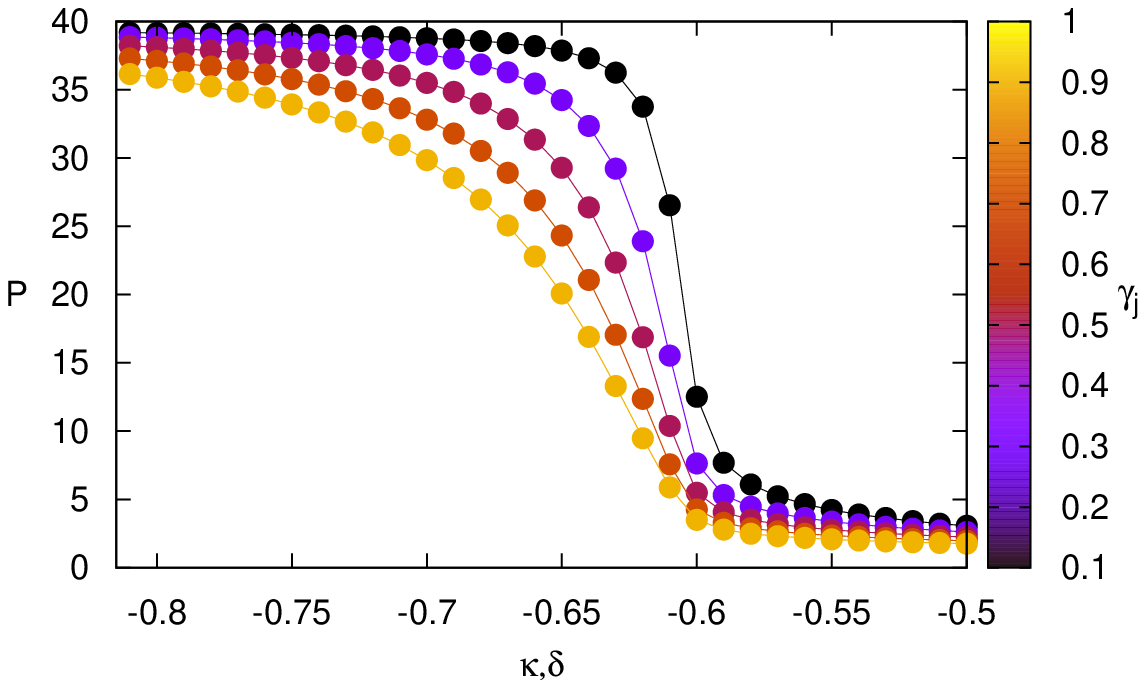,width=0.5\linewidth}
\caption{Values for participation number $P$ for various values of coupling strength $\delta=\kappa$ and initial perturbation $\gamma_j$, clearly showing the transition between Localised and de-Localised Behaviour.}\label{fig:pvals}
\end{figure}

Values for the participation number $P$ versus coupling strength $\delta=\kappa$ for various $\gamma_j$ are shown in Figure \ref{fig:pvals}, where $\gamma_j$ is the perturbed site value and all other $\gamma_n=0$.
A rapid increase in $P$ on increasing the coupling strength demonstrates a de-localisation transition that is dependent on $\gamma_j$.
As $\gamma_j\rightarrow0$ the transition becomes sharper around the value $\delta,\kappa\approx-0.6$, in agreement with Equation (\ref{eqn:bif}).
A similar coupling-dependent transition between local and large-scale patterns is also investigated in the next section for the domains formed in 2D lattices.

\subsection{Clustering and Coarsening in the 2D Lattices}\label{sec:clusters}

For measuring the level of coarsening we use the \emph{Mix-Norm} $N$, introduced in \cite{mathew2005multiscale} as a measure of mixing (and therefore conversely segregation).
The Mix-Norm is defined by:
\begin{equation}
N^2=\sum_{k,l}\frac{|a_{k,l}|^2}{\sqrt{1+k^2+l^2}},
\end{equation}
where $a_{k,l}$ are the coefficients of the Fourier transform of the field under investigation.
Therefore smaller numbers for $N$ indicate a more mixed, fine-scaled structure to the data, and vice-versa.

\begin{figure}
\centering
\epsfig{file=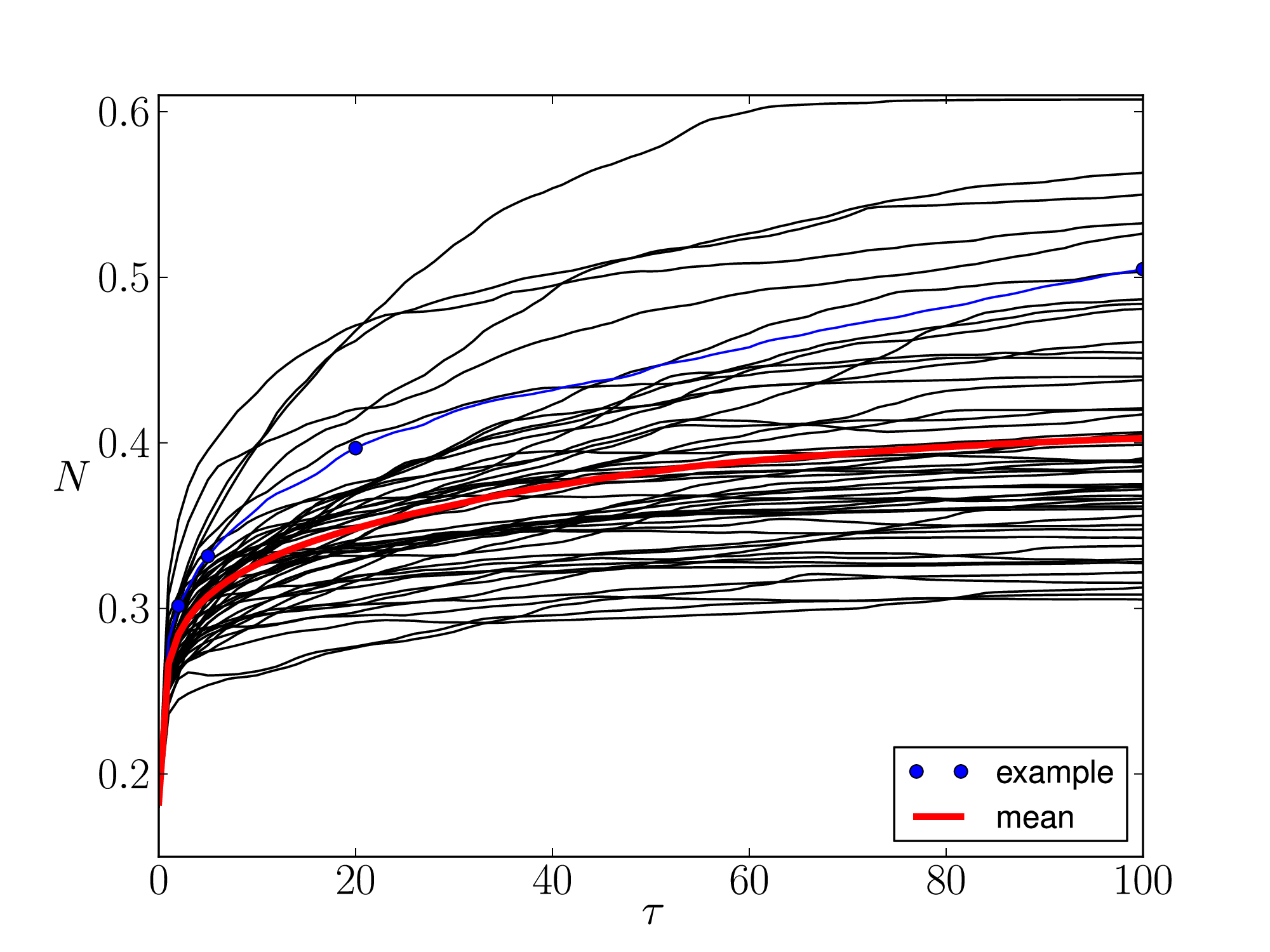,width=0.5\linewidth}
\caption{The Mix-Norm $N$, measuring de-mixing of random field of $x_0=\gamma_j$ on a 2D $N,M=50$ lattice with $\delta,\kappa=2$. 50 different randomizations are shown by thin (black) lines, demonstrating a wide variation depending on the initial condition. The specific example from Figure \ref{fig:results3} (d) is shown by (blue) circles.
The mean over all 50 runs is shown with the thick (red) line.}\label{fig:mixnorm}
\end{figure}

The development of $N$ for different random initial conditions (distributions of $-0.3<\gamma_i<0.3$) is given in Figure \ref{fig:mixnorm}.
These show different coarsening curves, with the final value of $N$ depending strongly on the precise micro-structure of the distributions.
However, all show a distinct coarsening Behaviour and investigations (not shown) revealed that the clusters become fixed in finite time, approaching a final value in $\tau$ of order 100.

\begin{figure}
\centering
\epsfig{file=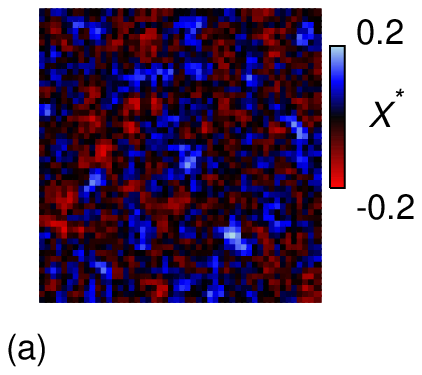,width=0.24\linewidth}
\epsfig{file=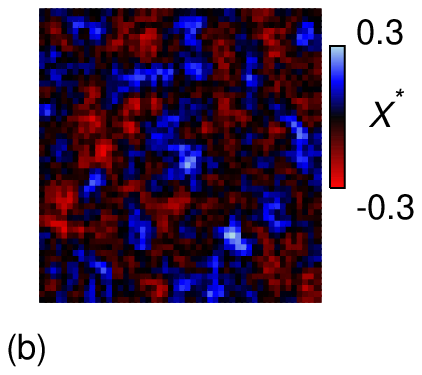,width=0.24\linewidth}
\epsfig{file=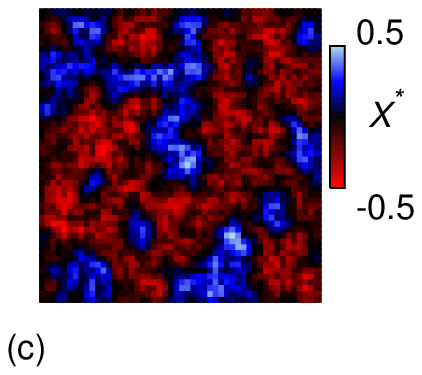,width=0.24\linewidth}
\epsfig{file=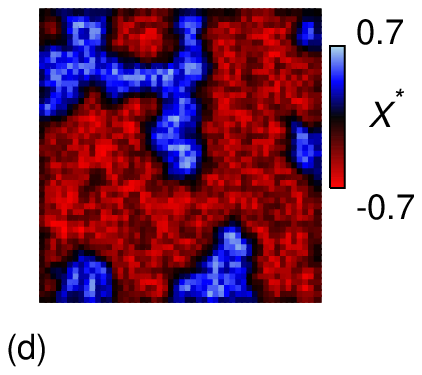,width=0.24\linewidth}

\caption{Examples of long-time ($\tau=100$) coarsening Behaviour at coupling strengths $\kappa,\delta=0.25,0.3,0.35,0.4$.  The same initial condition was used in each case and qualitatively different types of Behaviour can be observed.}\label{example4}
\end{figure}

\begin{figure}
\epsfig{file=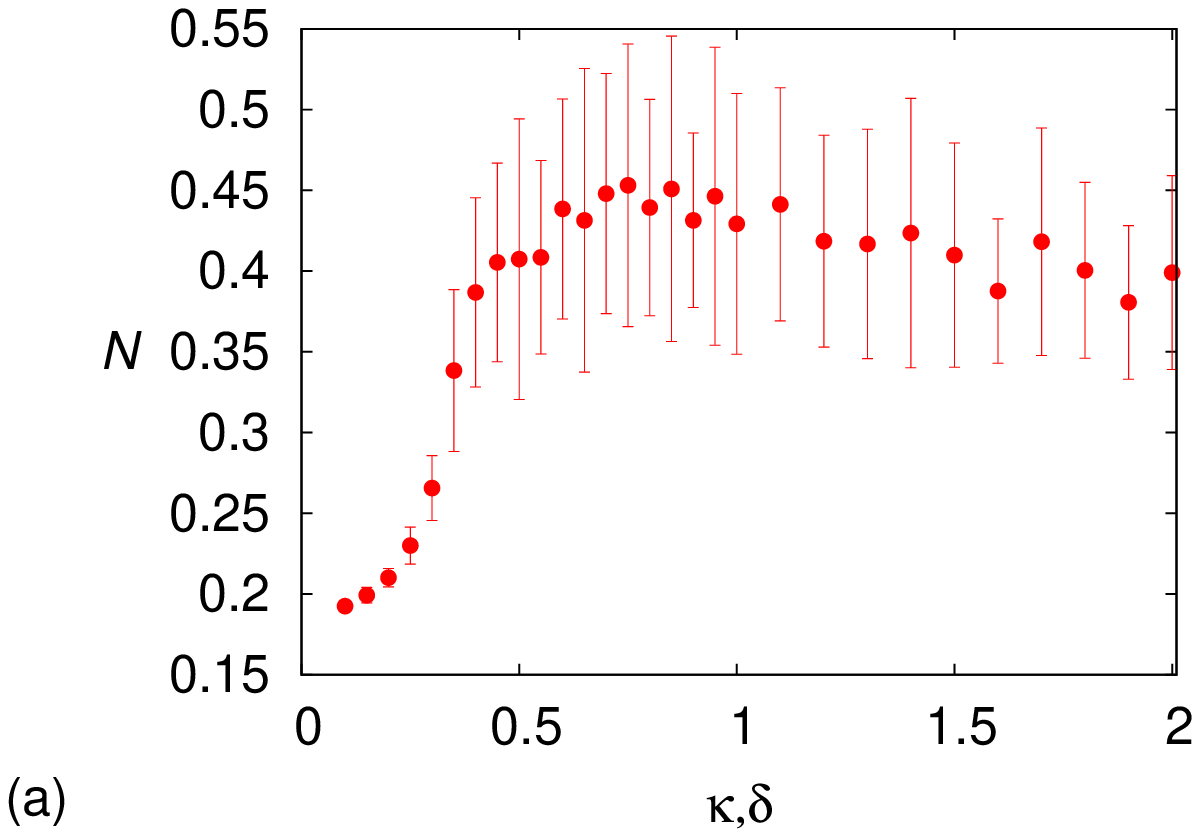,width=0.45\linewidth}
\epsfig{file=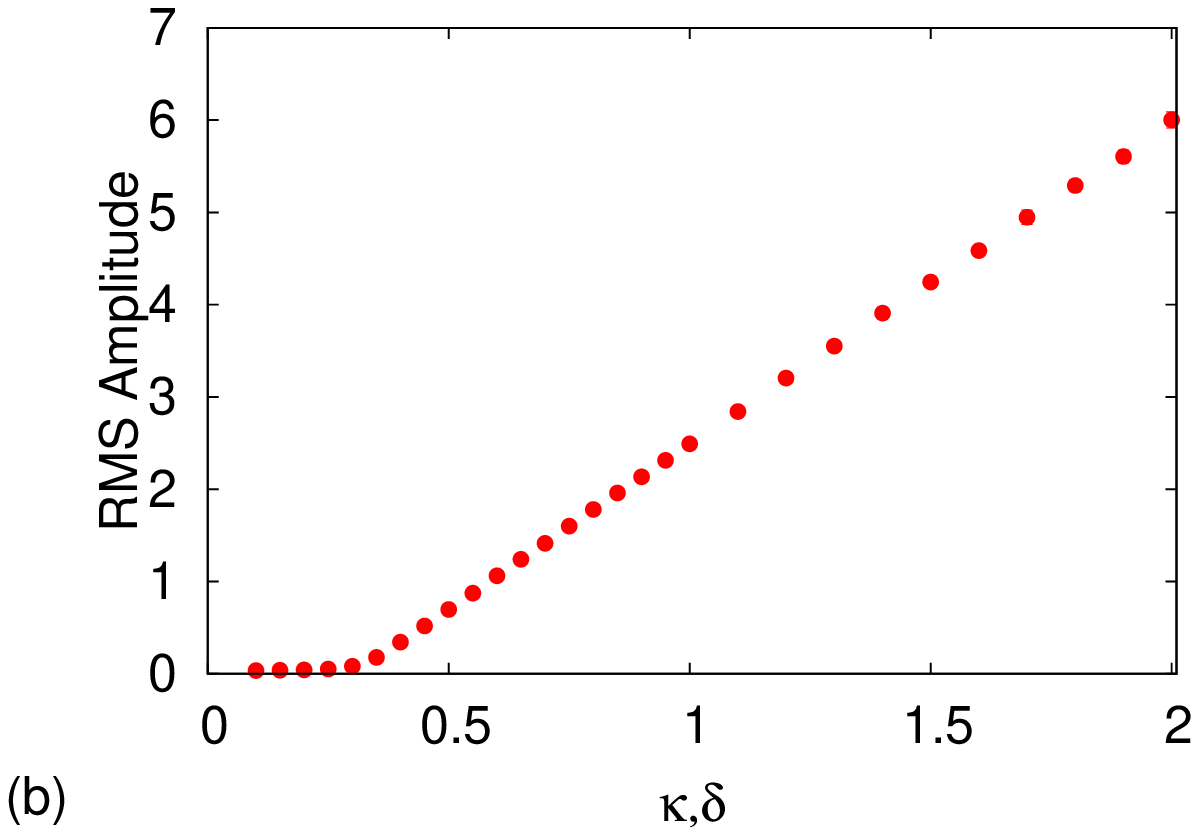,width=0.45\linewidth}
\caption{Average $N$ against $\kappa,\delta$ on both a linear (a) and semi-logarithmic (b) scale; showing the transition from local small-scale clustering to coarsened de-Localised patterns.}\label{fig:mixVdelta}
\end{figure}

The mean development of many distributions looks more regular, as also shown in Figure \ref{fig:mixnorm}.
This allowed investigation of the dependence of coarsening on the strength of interaction between neighbours $\delta=\kappa$.
A transition between a long-time state with local small clusters and a large-scale coarsened pattern were seen above $\delta,\kappa\approx0.3$, agreeing with the analytical estimates given in Section \ref{sec:homog}.
Examples around the transition values of coupling strength are shown in Figure \ref{example4} at $\tau=100$.
For the detailed study of the transition 50 such calculations were performed for each $\delta=\kappa$ and the mean of the Mix-Norm $N$ at $\tau=100$ is plotted against these coupling values in Figure \ref{fig:mixVdelta} (a).
In Figure \ref{fig:mixVdelta}(b), the RMS amplitude of the field is shown; further highlighting a critical transition in the Behaviour.
A clear transition can indeed be seen above a critical coupling strength, with the clustering measure growing by nearly two orders of magnitude.

\section{Interpretation and Conclusions}

Detailed analytical and numerical studies into the dynamics of networks of interacting individuals have been described in this paper.
These have shown good mutual agreement and and provided a wealth of results on aspects of cooperative Behaviour.
The results presented herein are interesting in their own right, with the results relating to clustering and group Behaviour, with its dependence on the strength of interactions, being potentially applicable to a range of similar systems.
Importantly here, in the context of the original problem they provide insight into the possible emergent Behaviour that could be encountered.
The results can be interpreted in terms of a market of interacting consumers who have to make decisions about whether or not to buy some product, here in the context of energy markets.

Investigating a range of coupling values and distributions of initial conditions has allowed us to understand the qualitative dynamics of the system.
For simplicity, homogeneous coupling was considered, along with uniform response Behaviour of the model consumers; allowing a clear connection to be made to the mathematical analysis.
The choice of coupling was found to lead to various qualitatively different regimes.
In the case of `antagonistic' coupling ($\delta,\kappa<0$) the development of spatial instabilities prevailed for different coupling strengths. 
This resulted in the initial spatial structure being destroyed in favour of a quasi-homogeneous regime; characterized by the alternation of the sign of activity in the neighbours (a checkerboard pattern). 
The interpretation for the applied problem would be that negative information exchange, such as mistrust or misinformation between individuals, generally leads to the loss of the reasonable decision-making among consumers.
This would result in destruction of the initial pattern of opinions and complete destabilization of the network, not leading to any clear consensus.
In this case the localisation of a perturbation, such as a strong-willed individual, was studied as a function of the strength of interaction, both analytically and numerically.
It was found that a transition between localisation of the disturbance and system-wide instability is critically dependent on the coupling magnitude, with the values in agreement between theory and simulation.

In cases where the information exchanged was cooperative, in the sense of it encouraging similar Behaviour amongst peers (with coupling coefficients taking positive values), the qualitative nature of the final state also depended strongly on the magnitude of coupling.
In the real-life situation this would be interpreted as the relative value placed on peers' opinions as compared to ones own bias.
Weak coupling slightly smoothed the variation of the initial opinion strengths across neighbours, resulting in weak spatial clusters which did not group together over time. 
Strong coupling was found to make the in-cluster distribution considerably more homogeneous, and the inter-cluster boundaries get sharper. 
In addition the clusters merge over time, resulting in a coarsening of the pattern.
This can be interpreted as herding of opinions in the network, finally leading to a one or two groups of consumers coming to a collective decision.
Again, the dependence of the transition from small (local) groups of opinions to de-Localised, large-scale patterns of Behaviour on the strength of interactions was investigated.
The results again revealed that there was a clear transition point, which can be interpreted as there being a minimum level of information exchange required to effect a consensus decision in the market of consumers.

Finally, this investigation has used various simplifying assumptions to ease comprehension and make the connection to analytical results more clear.
Future studies can investigate situations where these assumptions are relaxed to represent the real situation more naturally.
For example the network topology in real social groups is not as regular as considered here.
While people do interact with their next-door neighbours, there are also longer-range connections through work, family and other social ties.
Therefore a more realistic model might be a variation of the Watts--Strogatz (semi) random networks, which have both these features \cite{watts2003small}.
In addition, the interactions between and opinion-forming responses of individuals would not be homogeneous in the real world, so making these non-uniform will be a useful next step in dynamically modelling the decision-making Behaviour of consumer networks.

\section*{Acknowledgments}

We gratefully acknowledge the support of the UK EPSRC in funding this work under the ``Energy Challenges for Complexity Science'' program, for the project \emph{``Future Energy Decision Making for Cities - Can Complexity Science Rise to the Challenge?'' (EP/G059780/1)}.

\bibliography{ref_cell}
\bibliographystyle{ws-ijbc}
 
\end{document}